\documentclass[APA,Times1COL]{WileyNJDv5} 

\usepackage{amsthm,amsmath,amsfonts,amssymb}
\usepackage[authoryear]{natbib}
\usepackage{hyperref}
\usepackage{graphicx}
\usepackage{bm}
\usepackage{multicol}
\usepackage{multirow}

\newcommand{\bfe}{\bm{e}}

\newcommand{\bfh}{\bm{h}}

\newcommand{\bft}{\bm{t}}

\newcommand{\bfz}{\bm{z}}

\newcommand{\bfgamma}{\bm{\gamma}}

\newcommand{\bfeta}{\bm{\eta}}

\newcommand{\bfvarpi}{\bm{\varpi}}

\newcommand{\bfSigma}{\bm{\Sigma}}

\newcommand{\bfA}{\bm{A}}

\newcommand{\bfC}{\bm{C}}
\newcommand{\bfD}{\bm{D}}
\newcommand{\bfE}{\bm{E}}

\newcommand{\bfQ}{\bm{Q}}
\newcommand{\bfR}{\bm{R}}
\newcommand{\bfS}{\bm{S}}
\newcommand{\bfT}{\bm{T}}
\newcommand{\bfU}{\bm{U}}

\newcommand{\bfW}{\bm{W}}
\newcommand{\bfX}{\bm{X}}
\newcommand{\bfY}{\bm{Y}}
\newcommand{\bfZ}{\bm{Z}}

\newcommand{\bbI}{\mathbb{I}}
\newcommand{\bbP}{\mathbb{P}}

\newcommand{\calA}{\mathcal{A}}
\newcommand{\calB}{\mathcal{B}}
\newcommand{\calC}{\mathcal{C}}

\newcommand{\calG}{\mathcal{G}}

\newcommand{\calN}{\mathcal{N}}

\newcommand{\calP}{\mathcal{P}}

\newcommand{\calU}{\mathcal{U}}

\newcommand{\sumi}{\sum_{i=1}^I}
\newcommand{\sumj}{\sum_{j=1}^J}

\newcommand{\sumk}{\sum_{k=1}^{N_{ij}}}
\newcommand{\suma}{\sum_{a\in\calA}}

\articletype{Article Type}%

\received{Date Month Year}
\revised{Date Month Year}
\accepted{Date Month Year}
\journal{Journal}
\volume{00}
\copyyear{2024}
\startpage{1}

\begin{document}

\title{Model-assisted analysis of covariance estimators for stepped wedge cluster randomized experiments}

\author[1]{Xinyuan Chen}
\author[2,3]{Fan Li}

\authormark{CHEN and LI}
\titlemark{MODEL-ASSISTED STEPPED WEDGE CLUSTER RANDOMIZED EXPERIMENTS}

\address[1]{\orgdiv{Department of Mathematics and Statistics}, \orgname{Mississippi State Universit}, \orgaddress{\state{Mississippi State, MS}, \country{USA}}}

\address[2]{\orgdiv{Department of Biostatistics}, \orgname{Yale School of Public Health}, \orgaddress{\state{New Haven, CT}, \country{USA}}}

\address[3]{\orgdiv{Center for Methods in Implementation and Preventive Science}, \orgname{Yale School of Public Health}, \orgaddress{\state{New Haven, CT}, \country{USA}}}

\corres{Xinyuan Chen, Department of Mathematics and Statistics, Mississippi State University, Mississippi State, MS, 39762, USA. \email{xchen@math.msstate.edu} \\
	Fan Li, Department of Biostatistics, Yale School of Public Health, New Haven, CT, 06510, USA. \email{fan.f.li@yale.edu}}


\abstract[Abstract]{Stepped wedge cluster randomized experiments (SW-CREs) represent a class of unidirectional crossover designs. Although SW-CREs have become popular, definitions of estimands and robust methods to target estimands under the potential outcomes framework remain insufficient. To address this gap, we describe a class of estimands that explicitly acknowledge the multilevel data structure in SW-CREs and highlight three typical members of the estimand class that are interpretable. We then introduce four analysis of covariance (ANCOVA) working models to achieve estimand-aligned analyses with covariate adjustment. Each ANCOVA estimator is model-assisted, as its point estimator is consistent even when the working model is misspecified. Under the stepped wedge randomization scheme, we establish the finite population Central Limit Theorem for each estimator. We study the finite-sample operating characteristics of the ANCOVA estimators in simulations and illustrate their application by analyzing the Washington State Expedited Partner Therapy study.}


\jnlcitation{\cname{%
\author{Chen X.}, and
\author{Li F}}.
\ctitle{Model-assisted analysis of covariance estimators for stepped wedge cluster randomized experiments.} \cjournal{\it Scandinavian Journal of Statistics.} \cvol{2024;00(00):1--24}.}

\maketitle

\section{Introduction} \label{sec:intro}

Stepped wedge cluster randomized experiments (SW-CREs), alternatively referred to as stepped wedge designs, are frequently used to assess the causal effects of candidate treatments in public health, medicine, and implementation science research and are increasingly popular in pragmatic clinical trials. Under a stepped wedge design, all clusters are recruited at baseline and placed under the usual care condition; the intervention will then be rolled out across the follow-up periods in a staggered fashion until all clusters are exposed under the intervention \citep{Hussey2007}. In other words, each cluster will be randomized to a specific time point when the intervention starts to roll out and can be mapped to a monotonic treatment sequence over multiple periods. Compared to the conventional parallel-arm cluster randomized designs, stepped wedge designs are particularly attractive when the concurrent implementation of the candidate treatment may incur substantial stress on administrative planning and logistical infrastructure or when there is a desire to ensure the complete rollout of the intervention in all clusters during the course of the study. \citet{Hemming2020} provided four broad justifications for when the stepped wedge design is appropriate. Over the past decade, research on stepped wedge designs has placed much emphasis on trial planning to achieve sufficient statistical power with different intracluster correlation structures; a review of software for planning SW-CREs can be found in \citet{ouyang2022sample}. Although individual randomization is possible under a stepped wedge design, it is rare in practice; hence, we focus on cluster randomization as a more typical assignment mechanism for stepped wedge designs.

Despite the numerous efforts devoted to study planning, robust methods for analyzing SW-CREs have received relatively less attention, with a few exceptions such as \citet{thompson2018robust,hughes2020robust,kenny2021analysis} and \citet{maleyeff2022assessing}. Two main challenges remain not fully addressed in these previous studies. First, the typical analysis strategies include mixed-effects regression and generalized estimating equations with multilevel working correlation structures but do not explicitly address whether the treatment effect estimates were intended to generalize to the expected value of outcomes when applied to new cluster populations or new individual populations. In parallel-arm cluster randomized experiments, the use of model-based methods has been shown to produce treatment effect estimates that are not always straightforward to interpret \citep{wang2022two}, particularly when there is treatment effect heterogeneity according to cluster size, or equivalently, when the cluster size is informative. As the targets of inference, estimands are ideally defined at the outset to facilitate decision-making, and a clear description of estimands is vitally important for cluster randomized experiments, as exemplified by \citet{kahan2022estimands,kahan2023demystifying} and \citet{wang2022model}. Second, with clearly defined target estimands, analytical methods should ideally be model-assisted rather than model-based in that the estimated treatment effects are unbiased for the specified estimands even if certain model assumptions fail to hold and statistically efficient so that the uncertainty around the effect estimates is minimal to ensure a greater chance of identifying the treatment effect when it exists. As the resources for carrying out randomized experiments are not unlimited, methods with higher efficiency are critical for improving evidence generation through stepped wedge designs, and leveraging baseline covariates is a promising technique. However, recent methodological reviews \citep{Li2021,li2022stepped} devoted to SW-CREs showed that (1) model-based methods with parametric assumptions are prevalent, (2) estimands are not defined at the outset but rather considered as the regression coefficients of the treatment variable under the mixed-effects modeling or generalized estimating equations framework, and (3) few methodological developments considered incorporating baseline covariates (an exception is found in \citet{li2024planning}). Therefore, the extent to which covariate adjustment can improve efficiency without compromising clearly defined estimands in analyzing stepped wedge designs requires important clarification.

In this article, we adopt a finite population perspective to define relevant causal estimands that acknowledge the multilevel data structure of a stepped wedge design and to study robust regression-adjusted estimators for identifying these estimands. Our estimands generalize the counterparts in parallel-arm designs considered in \citet{Su2021} and \citet{kahan2022estimands} to longitudinal cluster randomized designs. Apart from defining different versions of causal estimands, we also provide different formulations of analysis of covariance (ANCOVA) models that enable model-assisted estimation. The ANCOVA working models differ with respect to considerations of period-specific covariate main effects and treatment-by-covariate interactions. We establish, for each ANCOVA estimator, the finite population Central Limit Theorem \citep{Li2017}, and motivate a design-based standard error estimator \citep{Scott1981,Schochet2021} under the staggered rollout randomization scheme. We conduct simulation studies to evaluate the finite-sample properties of the proposed estimators and compare their performances in terms of estimation efficiency. Our simulation results substantiate the large-sample consistency and asymptotic normality of proposed estimators and confirm that estimators adjusting for covariates lead to improved efficiency over the unadjusted estimators; however, the optimal estimator can depend on the data-generating process. In addition, we also compare the performance of the design-based standard error estimator with the conventional cluster-robust standard error \citep{Liang1986} (originally developed under a super population perspective for generalized estimating equations) through simulations to generate practical recommendations. Specifically, our empirical results point to a consistent message---if a sufficient number of clusters are present, we recommend ANCOVA estimators with treatment-by-covariate interactions, but if only a limited number of clusters are present, the ANCOVA estimator with treatment-by-covariate interactions but without period-specific covariate main effects may be preferred for its finite-sample stability. The choice of standard error estimators is further discussed in Section \ref{sec:discussion}.

Our work contributes to the burgeoning literature on design-based analysis of cluster randomized experiments. For parallel-arm cluster randomized designs, \citet{Imai2009} and \citet{Middleton2015} discussed nonparametric difference-in-means estimators for the average treatment effect in the absence of covariates. \citet{Schochet2021} established the finite population Central Limit Theorem of an ANCOVA estimator applied to blocked cluster randomization and compared the operating characteristics of the design-based standard error estimator and the cluster robust sandwich standard error estimator analytically and via simulations. \citet{Su2021} extended their results to cluster randomized experiments and elucidated the efficiency implications due to covariate adjustment from ANCOVA estimators under parallel-arm randomization. Our work differs from these existing works due to its focus on SW-CREs with staggered treatment rollout---a unique feature that necessitates novel developments for model-assisted inference. First, the staggered treatment rollout in SW-CREs leads to additional variations in the treatment effect estimands, as compared to those developed in parallel-arm cluster randomized experiments in \citet{kahan2022estimands,kahan2023demystifying} and \citet{Su2021}. Second, as we explain in Section \ref{sec:theory}, the staggered treatment rollout requires a more careful formulation of the potential outcomes that can depend on the adoption date and dictates the randomization-based distribution of the treatment effect estimators. This is a central difference from the simpler assignment mechanism considered in \citet{Schochet2021} and \citet{Su2021}, and has motivated us to obtain the asymptotic results from a multiple treatments perspective. On the other hand, our work is also connected to the literature on design-based difference-in-differences; see, for example, \citet{Athey2022, Callaway2021, deChaisemartin2020, Roth2021, Schochet2022, Sun2021}. However, these efforts were primarily focused on unit-level treatment assignments without the multilevel data structure, whereas we specifically focused on cluster-level randomization and individual-level data analysis. We also focused on establishing the finite population Central Limit Theorem under the Stable Unit Treatment Value Assumption (SUTVA) where there is a single version of intervention \citep{Hussey2007,Li2021} agnostic to the treatment duration (i.e. no learning effect or weakening effect over time if applied to the same unit). This assumption is commonly invoked to analyze SW-CREs in the literature, and under this assumption, we seek to clarify appropriate causal estimands and survey a class of model-assisted analytical strategies.

The rest of this article is structured as follows. Section \ref{sec:estimand} introduces the finite population causal inference framework and discusses causal estimands for SW-CREs. Section \ref{sec:ancova} introduces four ANCOVA estimators and establishes their theoretical properties under finite population asymptotic regimes. Section \ref{sec:simulation} and \ref{sec:application} present results from our simulation study and the illustrative analysis of the Washington State Expedited Partner Therapy study. Section \ref{sec:discussion} concludes with a discussion and outlines directions for future research.

\section{Notation and Estimands} \label{sec:estimand}
\subsection{Assumptions for stepped wedge designs}

We consider the cross-sectional stepped wedge design, where different individuals are included in each cluster at the beginning of each distinct period. This is the most common design in practice when clusters are rolled out in a staggered fashion, according to a recent systematic review by \citet{nevins2023adherence}. In a standard stepped wedge design, there are three experimental phases: the pre-rollout, where no clusters receive treatment (i.e., all clusters are placed under the control condition); the rollout, where clusters are randomized to different treatment schedules in a staggered fashion, and the post-rollout, where all clusters have received the treatment. To formalize a standard stepped wedge design (standard in the sense that there is a single pre-rollout period, a single post-rollout period, and at least one new cluster receiving treatment in each rollout period), we assume a study with $I$ clusters (indexed by $i$) and $J+2$ periods (indexed by $j$), where period 0 is the pre-rollout, periods $\{1,\ldots,J\}$ are the rollout, and period $J+1$ is the post-rollout. In general, the length of each period is pre-determined in the study planning stage, and each period, often of equal length, is associated with at least one new cluster switching to treatment; see Figure \ref{fig:stepped} for a schematic illustration of the standard stepped wedge design. We use $Z_{ij}\in\{0,1\}$ to denote the treatment status indicator for cluster $i\in\{1,\ldots,I\}$ in period $j\in\{0,1,\ldots,J,J+1\}$, and the number of treated clusters in each rollout period $j$, $I_j$, is known and fixed for all $j$. The rollout starts with $I_1$ clusters randomized to treatment in period 1, and in period 2, the previously selected $I_1$ clusters remain treated, while $I_2-I_1$ out of the $I-I_1$ untreated clusters are randomized to treatment, and so forth. This process ends with all clusters being eventually treated in period $J+1$. It is immediate that $Z_{i0}=0$ and $Z_{i,J+1}=1$ due to the definition of the pre-rollout and post-rollout period, and that $0=I_0<I_1< I_2<\cdots< I_J<I_{J+1}=I$. This specific randomization design implies that elements of the treatment sequence vector $\bfZ_i=(Z_{i0},Z_{i1},\ldots,Z_{ij},Z_{i,J+1})'$ are correlated, with $Z_{i0}=0$, $Z_{i,J+1}=1$, and $Z_{ij'}=1$ if $Z_{ij}=1$ for all $j'>j$. Additionally, it can be shown that the marginal distribution of the treatment variable across clusters in each period $j$, $(Z_{1j},\ldots,Z_{Ij})$, follows a hyper-geometric distribution with parameters $(I,I_j)$. 

\begin{figure}[htbp]
	\setlength{\unitlength}{0.14in} 
	\centering 
	\begin{picture}(31.5,11) 
		\setlength\fboxsep{0pt}
		\put(1,8){\framebox(5.6,1.5){\footnotesize$\{N_{10},\ldots,N_{I_1,0}\}$}}
		\put(8.5,8){\colorbox{blue!45}{\framebox(5.6,1.5){\footnotesize$\{N_{11},\ldots,N_{I_1,1}\}$}}}
		\put(16,8){\colorbox{blue!45}{\framebox(5.6,1.5){\footnotesize$\{N_{12},\ldots,N_{I_1,2}\}$}}}
		\put(23.5,8){\colorbox{blue!45}{\framebox(5.6,1.5){\footnotesize$\{N_{13},\ldots,N_{I_1,3}\}$}}}
		\put(31,8){\colorbox{blue!45}{\framebox(5.6,1.5){\footnotesize$\{N_{14},\ldots,N_{I_1,4}\}$}}}
		\put(1,6){\framebox(5.6,1.5){\footnotesize$\{N_{I_1+1,0},\ldots,N_{I_2,0}\}$}}
		\put(8.5,6){\framebox(5.6,1.5){\footnotesize$\{N_{I_1+1,1},\ldots,N_{I_2,1}\}$}}
		\put(16,6){\colorbox{blue!45}{\framebox(5.6,1.5){\footnotesize$\{N_{I_1+1,2},\ldots,N_{I_2,2}\}$}}}
		\put(23.5,6){\colorbox{blue!45}{\framebox(5.6,1.5){\footnotesize$\{N_{I_1+1,3},\ldots,N_{I_2,3}\}$}}}
		\put(31,6){\colorbox{blue!45}{\framebox(5.6,1.5){\footnotesize$\{N_{I_1+1,4},\ldots,N_{I_2,4}\}$}}}
		\put(1,4){\framebox(5.6,1.5){\footnotesize$\{N_{I_2+1,0},\ldots,N_{I_3,0}\}$}}
		\put(8.5,4){\framebox(5.6,1.5){\footnotesize$\{N_{I_2+1,1},\ldots,N_{I_3,1}\}$}}
		\put(16,4){\framebox(5.6,1.5){\footnotesize$\{N_{I_2+1,2},\ldots,N_{I_3,2}\}$}}
		\put(23.5,4){\colorbox{blue!45}{\framebox(5.6,1.5){\footnotesize$\{N_{I_2+1,3},\ldots,N_{I_3,3}\}$}}}
		\put(31,4){\colorbox{blue!45}{\framebox(5.6,1.5){\footnotesize$\{N_{I_2+1,4},\ldots,N_{I_3,4}\}$}}}
		\put(1,2){\framebox(5.6,1.5){\footnotesize$\{N_{I_3+1,0},\ldots,N_{I_4,0}\}$}}
		\put(8.5,2){\framebox(5.6,1.5){\footnotesize$\{N_{I_3+1,1},\ldots,N_{I_4,1}\}$}}
		\put(16,2){\framebox(5.6,1.5){\footnotesize$\{N_{I_3+1,2},\ldots,N_{I_4,2}\}$}}
		\put(23.5,2){\framebox(5.6,1.5){\footnotesize$\{N_{I_3+1,3},\ldots,N_{I_4,3}\}$}}
		\put(31,2){\colorbox{blue!45}{\framebox(5.6,1.5){\footnotesize$\{N_{I_3+1,4},\ldots,N_{I_4,4}\}$}}}
		\put(1,1.5){$\underbrace{\hspace{5.7em}}_{pre-rollout}$}
		\put(8.5,1.5){$\underbrace{\hspace{20.9em}}_{rollout}$}
		\put(31,1.5){$\underbrace{\hspace{5.7em}}_{post-rollout}$}
		\put(-6.8,2.5){$\{I_3+1,\ldots,I_4\}$}\put(-6.8,4.5){$\{I_2+1,\ldots,I_3\}$}\put(-6.8,6.5){$\{I_1+1,\ldots,I_2\}$}
		\put(-5.1,8.5){$\{1,\ldots,I_1\}$}
		\put(-5.3,10.5){Clusters}\put(2,10.5){Period $0$}\put(9.5,10.5){Period $1$}
		\put(17,10.5){Period $2$}\put(24.5,10.5){Period $3$}\put(32,10.5){Period $4$}
	\end{picture}
	\caption{An example schematic of a standard SW-CRE with $I=I_4$ clusters and $J=3$ rollout periods. There are a total of $4$ possible treatment adoption time points, and a subset of all clusters, represented by each row, has a unique treatment adoption time. Each period is of equal length, and we include in each cell all cluster-period sizes ($N_{ij}$) for the clusters included in that row during that period. A white cell indicates the control condition and a blue-shaded cell indicates the treatment condition. In this standard design, Period 0 and Period 4 are the pre-rollout and post-rollout periods, and the middle periods are the rollout periods. Of note, $I_1< I_2< I_3< I_4$, and each period is associated with at least one new cluster switching from control to treatment.}
	\vspace*{-0.1in}
	\label{fig:stepped}
\end{figure}

Under a cross-sectional design, different individuals are included during each period within each cluster, and the cluster-period sizes, denoted by $N_{ij}$, are usually all different. This is the case, for example, when hospitals (clusters) are randomized and $N_{ij}$ represents the number of individuals seeking healthcare in cluster-period $(i,j)$ or when geographical regions (clusters) are randomized and $N_{ij}$ represents the size of the distinct subpopulation included in cluster-period $(i,j)$ for outcome measurement; our data example in Section \ref{sec:application} follows the latter scenario. For individuals $k=1,\ldots,N_{ij}$ in cluster-period $(i,j)$, we proceed to define the potential outcomes. We let $A_i=a\in\calA=\{1,\ldots,J,J+1\}$ denote the period index such that cluster $i$ first receives treatment (the so-called treatment adoption time), and therefore $Z_{ij}=\bbI\{A_i\leq j\}$. Without any further assumptions, each individual has potential outcome $Y_{ijk}^{A_i,\bfA_{-i}}$ that depends not only on the adoption time for cluster $i$ but also those for other clusters ($\bfA_{-i}$ is the vector of adoption time for all other clusters). We first impose a cluster-level SUTVA, formalized in Assumption \ref{asp:sutva}. 

\begin{assumption}[Cluster-level SUTVA] \label{asp:sutva}
	Let $Y_{ijk}^{A_i,\bfA_{-i}}$ denote the potential outcome for an individual given the adoption time of all clusters, then (i) $Y_{ijk}^{A_i,\bfA_{-i}} = Y_{ijk}^{A_i,\bfA_{-i}^*}$, $\forall~\bfA_{-i}\neq \bfA_{-i}^*$; and (ii) $Y_{ijk}^{A_i,\bfA_{-i}} = Y_{ijk}^{A_i^*,\bfA_{-i}^*}$ if $A_i=A_i^*$.
\end{assumption}

Assumption \ref{asp:sutva} implies that individual-level potential outcomes from cluster $i$ in period $j$ only depend on the specific treatment assignment of cluster $i$ but not on assignments for other clusters. This is a conventional assumption made for cluster randomized experiments and is likely plausible if the clusters are not in close geographical proximity such that members from different clusters do not interact with each other. It rules out interference between clusters and allows us to define potential outcome $Y_{ijk}^{A_i}$ without ambiguity. However, the notation $Y_{ijk}^{A_i}$ implicitly assumes that the potential outcome can depend on the treatment adoption time and, therefore, the duration of the treatment. While this form is plausible in settings where there is a learning effect or a weakening effect over time \citep{hughes2015current,kenny2021analysis,maleyeff2022assessing}, in this article, we focus on a simpler setting and consider the following assumption to reduce the number of potential outcomes based on a dichotomy of treatment received. 

\begin{assumption}[Treatment duration irrelevance] \label{asp:dur}
	There is only one version of treatment across different periods, and variations in treatment duration are irrelevant to the potential outcomes. That is, for each $k\in\{1,\ldots,N_{ij}\}$, (i) $Y_{ijk}^a = Y_{ijk}^{a'} = Y_{ijk}(1), ~\text{if}~ \max\{a,a'\}\leq j$; (ii) $Y_{ijk}^a = Y_{ijk}^{a'} = Y_{ijk}(0), ~\text{if}~ \min\{a,a'\}>j$; (iii) $Y_{ijk}^a = Y_{ijk}(1), ~ Y_{ijk}^{a'} = Y_{ijk}(0), ~\text{if}~ a\leq j<a'$, for $a$, $a'\in\calA$ and $j\in\{0,1,\ldots,J,J+1\}$.
\end{assumption}

Assumption \ref{asp:dur} implies that individual-level potential outcomes from cluster $i$ in period $j$ only depend on the treatment received at that period, $Z_{ij}$, such that we can write $Y_{ijk}(Z_{ij})$ without ambiguity. Furthermore, Assumption \ref{asp:dur} can be considered a version of the treatment-variation irrelevance assumption  \citep{vanderweele2009concerning} to rule out multiple treatment versions. To this extent, Assumptions \ref{asp:sutva} and \ref{asp:dur} can be jointly considered as a generalized SUTVA applied to SW-CREs. Indeed, the treatment duration irrelevance has been conventionally assumed in the literature for stepped wedge designs \citep{Li2021}, even though rarely stated formally. It can also be considered a multilevel extension of the no anticipation and the invariance to history assumptions adopted in the difference-in-differences literature \citep{Athey2022}. Under Assumption \ref{asp:dur}, we can write the observed individual-level outcomes as $Y_{ijk}=Z_{ij}Y_{ijk}(1)+(1-Z_{ij})Y_{ijk}(0)$ for each rollout period $j$. However, the observed outcome $Y_{i0k}=Y_{i0k}(0)$ and $Y_{i,J+1,k}=Y_{i,J+1,k}(1)$ for all $k$ by definition of the pre-rollout and post-rollout periods. 

To complete the data specification, we assume the cluster-period size $N_{ij}$ to be unaffected by the treatment assignment; thus ruling out post-randomization selection bias \citep{Li2022_CT}. We write the number of individuals in cluster $i$ during rollout as $N_i=\sumj N_{ij}$, the number of individuals in period $j$ as $N_j=\sumi N_{ij}$, and the total number of individuals across clusters and all rollout periods as $N=\sumi\sumj N_{ij}$. We further denote $\bfQ_{ijk}$ as the vector of individual-level baseline covariates recorded during recruitment (assumed exogenous and not affected by the treatment assignment), $\bfC_{ij}$ the vector of cluster-level characteristics (can depend on period), possibly including components of cluster-level summaries, $\overline{\bfQ}_{ij}=N_{ij}^{-1}\sumk \bfQ_{ijk}$, summary of the pre-rollout outcomes, $\overline{Y}_{i0}=N_{i0}^{-1}\sum_{k=1}^{N_{i0}}Y_{i0k}$ and cluster-period size $N_{ij}$. The collection of covariates not affected by the treatment assignment for each individual can then be defined as $\bfX_{ijk}=\bfQ_{ijk}\cup\bfC_{ij}$. The observed data for each cluster-period, therefore, is $\{(Y_{ijk},Z_{ij},\bfX_{ijk}),k=1,\ldots,N_{ij}\}$. Finally, we state the randomization assumption.

\begin{assumption}[Stepped wedge randomization]\label{asp:rand}
	Write $\mathcal{Y}$ and $\mathcal{X}$ as the collection of all potential outcomes and covariates across individuals and cluster-periods, then 
	\begin{equation*}
		\bbP(\bfZ_i=\bfz|\mathcal{Y},\mathcal{X})={I\choose I_1,I_2-I_1,\ldots,I_{J+1}-I_J}^{-1},
	\end{equation*}
	where $Z_{i0}=0$ and $Z_{i1}=1$ almost surely.
\end{assumption}

Assumption \ref{asp:rand} defines the rollout schedule and states the source of randomness in the observed outcome. Importantly, we write $e_j=I_j/I$ as the cluster-level propensity score fixed by design and naturally have $0=e_0<e_1\leq e_2\leq\cdots\leq e_J<e_{J+1}=1$. Importantly, the cluster-level propensity score during the pre-rollout and post-rollout, $e_0=0$ and $e_{J+1}=1$, violating the positivity assumption \citep{Imbens2015}; this is a consequence of the particular study design, and because there is no possibility for individuals during these two periods (the pre-rollout and post-rollout periods) to receive the unobserved counterfactual treatment assignment, we notionally set $Y_{i0k}(1)=\star$ and $Y_{i,J+1,k}(0)=\star$, and only formulate estimands based on the rollout periods. The idea of considering the potential outcomes with no possibility to be observed as undefined values is reminiscent of the truncation-by-death problem in causal inference \citep{zhang2009likelihood}, and one can consider that $\{Y_{i0k}(1),Y_{i,J+1,k}(0)\}$ are truncated by the study design due to the inclusion of pre-rollout and post-rollout periods. We do acknowledge, however, that it might be possible to leverage additional assumptions to extrapolate $\{Y_{i0k}(1),Y_{i,J+1,k}(0)\}$ from the observed data; we do not pursue this idea here but provide a discussion on this point in Section \ref{sec:discussion}.

\subsection{Causal estimands}

We consider a finite population framework where all potential outcomes are fixed quantities, and the variability is solely driven by the randomization distribution. Under Assumptions \ref{asp:sutva} and \ref{asp:dur}, we are interested in the following class of weighted average treatment effect (WATE) estimands during the rollout periods, defined as
\begin{equation} \label{eq:gen-estimand}
	\begin{aligned}
		\tau = \sumj\frac{w_j}{\sumj w_j}\left[\frac{\sumi w_{ij}\left\{\overline{Y}_{ij}(1)-\overline{Y}_{ij}(0)\right\}}{\sumi w_{ij}}\right]=\overline{Y}(1)-\overline{Y}(0),
	\end{aligned}
\end{equation}
where the weighted cluster-period mean potential outcome is 
\begin{align*}
	\overline{Y}_{ij}(z)=\frac{\sumk w_{ijk}Y_{ijk}(z)}{\sumk w_{ijk}}, ~~\text{for } z\in\{0,1\},
\end{align*}
with individual-specific weight $w_{ijk}\geq 0$, cluster-period total weight, $w_{ij}=\sum_{k=1}^{N_{ij}} w_{ijk}$, and period total weight, $w_j=\sum_{i=1}^I w_{ij}$. Alternatively, \eqref{eq:gen-estimand} can be re-expressed as
\begin{equation*}
	\tau = \frac{\sumj\sumi\sumk w_{ijk}\left\{Y_{ijk}(1)-Y_{ijk}(0)\right\}}{\sumj\sumi\sumk w_{ijk}},
\end{equation*}
which is the weighted average of the individual-level treatment effects over individuals across all clusters and all rollout periods. This estimand has also been considered by \citet{Schochet2021} in blocked cluster randomized controlled experiments, and a subtle conceptual difference is that we have excluded the pre-rollout and post-rollout periods due to the absence of positivity. Furthermore, we observe that the weighted average treatment effect in period $j$ can be written as 
\begin{equation*}\label{eq:gen-estimand-p-j}
		\tau_j=\frac{\sumi w_{ij}\left\{\overline{Y}_{ij}(1)-\overline{Y}_{ij}(0)\right\}}{\sumi w_{ij}}=\overline{Y}_j(1)-\overline{Y}_j(0).
\end{equation*}
This quantity is a building block for the WATE estimand \eqref{eq:gen-estimand} in stepped wedge designs and is essentially a weighted average treatment effect in the spirit of \citet{Su2021} for a parallel-arm cluster randomized experiment.

Despite the generality of \eqref{eq:gen-estimand}, we focus on three specific members under different choices of the weights corresponding to interpretable estimands for SW-CREs. Firstly, the uniform weight is given by $w_{ijk}=1$ that weighs each individual equally. For this specification, the cluster-period total weight $w_{ij}=\sumk w_{ijk}=N_{ij}$, the period total weight $w_j=\sumi w_{ij}=N_j$, $\overline{Y}_{ij}(z)=\sumk Y_{ijk}(z)/N_{ij}$ and $\overline{Y}_{ij}=\sumk Y_{ijk}/N_{ij}$ become the simple average of the potential and observed outcomes in each cluster-period. The estimand in period $j$ is 
\begin{align}\label{eq:est-taup}
	\tau_j=
	\frac{\sumi\left\{N_{ij}\overline{Y}_{ij}(1)-N_{ij}\overline{Y}_{ij}(0)\right\}}{N_j}=
	\frac{\sumi\sumk \left\{Y_{ijk}(1)- Y_{ijk}(0)\right\}}{N_j},
\end{align}
which represents the individual-average treatment effect as defined in parallel-arm cluster randomized experiments \citep{kahan2022estimands}, as it is the average of individual-level counterfactual contrasts over individuals from all clusters in period $j$. The final estimand of the rollout population under the uniform weight is
\begin{align*}
	\tau^{ind}&
	=\frac{\sumj\sumi\sumk \left\{Y_{ijk}(1)- Y_{ijk}(0)\right\}}{\sumj N_j}=\overline{Y}^{ind}(1)-\overline{Y}^{ind}(0),
\end{align*}
which is the average of all individual-level counterfactual contrasts over individuals in all clusters and all rollout periods. Secondly, we consider the inverse period size weight, where $w_{ijk}=N_j^{-1}$. This specification resembles the scaled cluster total representation in \citet{Su2021} when analyzing parallel-arm cluster randomized experiments. In this case, $w_{ij}=\sumk w_{ijk}=N_j^{-1}N_{ij}$ and $w_j=\sumi w_{ij}=1$. Therefore, the estimand in period $j$ is equal to \eqref{eq:est-taup}. Since the period-specific total weight is $1$, the final estimand over the rollout population is given as
\begin{align*}
	\tau^{period}=\frac{\sumj\left\{N_j^{-1}\sumi\sumk \left\{Y_{ijk}(1)- Y_{ijk}(0)\right\}\right\} }{J}=\overline{Y}^{period}(1)-\overline{Y}^{period}(0),
\end{align*}
which is interpreted as the simple average of period-specific mean counterfactual contrasts over all rollout periods and referred to as the period-average treatment effect. Finally, we consider the inverse cluster-period size weight, where $w_{ijk}=N_{ij}^{-1}$. For this specification, we have $w_{ij}=\sumk w_{ijk}=1$, $w_j=\sumi w_{ij}=I$, $\overline{Y}_{ij}(z)=\sumk Y_{ijk}(z)/N_{ij}$ as the simple average of the potential outcomes in each cluster-period, leading to a period-specific estimand defined as $\tau_j=I^{-1}\sumi\left\{\overline{Y}_{ij}(1)-\overline{Y}_{ij}(0)\right\}$, which can be considered the cluster-average treatment effect or unit-average treatment effect defined for parallel-arm cluster randomized experiments \citep{Su2021,wang2022two}. The final causal estimand average across all rollout periods is therefore given by
\begin{align*}
	\tau^{cell}=\frac{\sumj\sumi\left\{\overline{Y}_{ij}(1)-\overline{Y}_{ij}(0)\right\}}{IJ} = \overline{Y}^{cell}(1)-\overline{Y}^{cell}(0),
\end{align*}
which is the average of all cluster-period-specific or cell-specific mean counterfactual contrasts. We refer to this estimand as the cell-average treatment effect, where each cell represents a unique cluster-period during the rollout. 

\begin{table}[htbp] 
	\caption{A summary of definitions for three interpretable causal estimands in the general family of estimands \eqref{eq:gen-estimand}}\label{tab:estimands}
	\vspace{-0.15in}
	\begin{center}
		\resizebox{\linewidth}{!}{
			\begin{tabular}{cll} 
				\hline 
				\textbf{Estimand} & \textbf{Choice of weight} & \multicolumn{1}{c}{\textbf{Interpretation}} \\
				\hline
				$\tau^{ind}=\overline{Y}^{ind}(1)-\overline{Y}^{ind}(0)$ & $w_{ijk}=1$ & The average of individual-level counterfactual contrasts,\\ 
				& $w_{ij}=N_{ij}$  & across all individuals during the rollout. This estimand gives \\
				& $w_j=N_j$ & equal weight to each individual.\\
				\hline 
				$\tau^{period}=\overline{Y}^{period}(1)-\overline{Y}^{period}(0)$ & $w_{ijk}=N_j^{-1}$ & The average of individual-level counterfactual contrasts per\\
				& $w_{ij}=N_{ij}/N_j$ & period, which is further averaged across all rollout periods. \\
				& $w_j=1$ & This estimand gives equal weight to each rollout period.\\
				\hline 
				$\tau^{cell}=\overline{Y}^{cell}(1)-\overline{Y}^{cell}(0)$ & $w_{ijk}=N_{ij}^{-1}$ & The average of cluster-period cell-level mean counterfactual\\ 
				& $w_{ij}=1$  & contrasts, across all cluster-period cells during rollout. This\\
				& $w_j=I$ & estimand gives equal weight to each cluster-period cell.\\
				\hline
			\end{tabular}
		}
	\end{center}
\end{table}

Table \ref{tab:estimands} provides a concise summary of the weight specifications and interpretations of these three estimands, which explicates whether the treatment effect estimates target the expected value of outcomes when applied to the population of all individuals, population of cluster-period cells, or populations of individuals in an average period. In a general setting where the cluster sizes are varying, these three estimands are not necessarily equal, especially when there is an association between the cluster-period size $N_{ij}$ and the within-cluster counterfactual contrasts; this has been referred to as the informative cluster size \citep{kahan2022estimands}. However, the three estimands will coincide when the cluster-period sizes $N_{ij}$'s are homogeneous or the treatment effect is constant across cluster-period cells. In the more general settings where the three estimands can take different values, they may be interpreted differently, and the choice of estimands will depend on the study context, the nature of the intervention, and the study objectives. Intuitively, the individual-average treatment effect answers the question ``{how effective is the intervention for an average individual during the entire rollout?}'', the period-average treatment effect answers the question ``{how effective is the intervention for an average individual during an average period?}'', whereas the cell-average treatment effect answers the question ``{how effective is the intervention for an average cluster-period cell?}'' Answers to these three questions pertain to different aspects of the intervention effect and may lead to different policy implications. The differentiation of estimands in SW-CREs in this article can be seen as the counterparts of those developed for parallel-arm cluster randomized experiments; see, for example, \citet{kahan2022estimands}.

\section{Point and variance estimation with analysis of covariance} \label{sec:ancova}

\subsection{Model formulation}

To estimate the class of estimands \eqref{eq:gen-estimand} and specific members in Table \ref{tab:estimands}, we first consider the following working ANCOVA model that allows for baseline covariate adjustment: 
\begin{equation}  \label{eq:gen-model}
	\begin{aligned}
		Y_{ijk} = \beta_{j} + \tau_{j}Z_{ij}+\widetilde{\bfX}_{ijk}\bfgamma+e_{ijk},
	\end{aligned}
\end{equation}
where $\beta_j$ is the period fixed effect or referred to as the secular trend parameter in the stepped wedge design literature, $\tau_j$ is the period-specific average treatment effect parameter, $\widetilde{\bfX}_{ijk}=\bfX_{ijk}-\overline{\bfX}_j$ is the $p$-dimensional period-mean centered baseline covariate row vector with $\overline{\bfX}_{ij}=\sumk w_{ijk}\bfX_{ijk}\allowbreak/\sumk w_{ijk}$ and $\overline{\bfX}_j=\sumi w_{ij}\overline{\bfX}_{ij}/\sumi w_{ij}$, $\bfgamma$ is the associated parameter vector, and $e_{ijk}$ is the individual-level random noise that is assumed to have a mean-zero distribution and independent across individuals. As will be seen in due course, ANCOVA model \eqref{eq:gen-model} is designed to ensure $\tau_j$ is connected with elements of our target estimands. Because no cluster-level fixed effects are included, \eqref{eq:gen-model} is only a working model and does not necessarily reflect the data-generating process. On the other hand, due to the absence of the cluster-level fixed effects, \eqref{eq:gen-model} is akin to the marginal (marginalizing over clusters) model in \citet{li2018sample} and \citet{tian2024information} for stepped wedge designs except that we include a period-specific treatment effect parameter and allow for covariate adjustment.

Model \eqref{eq:gen-model} has been used in \citet{Schochet2021} to analyze blocked cluster randomized experiments and shares some connections with the ANCOVA I estimator discussed in \citet{Lin2013} and \citet{Tsiatis2008} applied to individually randomized experiments. Therefore, we call model \eqref{eq:gen-model} the ANCOVA I model. Importantly, ANCOVA I is a working model in the sense that we can interpret $\tau_j$ as a causal effect parameter regardless of whether the linear index, $\widetilde{\bfX}_{ijk}\bfgamma$, is compatible with the unknown true data-generating process (DGP); for this reason, the resulting estimator for $\tau_j$ is model-assisted rather than model-based. The working ANCOVA I model also covers the unadjusted estimator where no covariate adjustments are included or by setting the parameter vector $\bfgamma=\bm{0}$.

Besides ANCOVA I, we also consider three possible variations of model formulation, depending on whether treatment-by-covariate or treatment-by-covariate-by-period interactions are considered in the working model. Specifically, the ANCOVA II model includes period-specific main effects of the covariates and is written as 
\begin{equation} \label{eq:gen-model-tv}
	\begin{aligned}
		Y_{ijk} = \beta_{j} + \tau_{j}Z_{ij}+\widetilde{\bfX}_{ijk}\bfgamma_{j}+e_{ijk}.
	\end{aligned}
\end{equation}
In the special case where $\bfgamma_{j}=\bfgamma$ for all $j$, the ANCOVA II model reduces to the ANCOVA I model. In the analyses of individually randomized and parallel-arm cluster randomized experiments, specifying the interactions between the treatment status and the covariates is often considered a strategy to further improve precision under unbalanced randomization \citep{Lin2013}. We thus further include the following ANCOVA III model with treatment-by-covariate interactions: 
\begin{equation} \label{eq:int-model}
	\begin{aligned}
		Y_{ijk} = \beta_{j} + \tau_{j}Z_{ij}+\widetilde{\bfX}_{ijk}\bfgamma+Z_{ij}\widetilde{\bfX}_{ijk}\bfeta +e_{ijk},
	\end{aligned}
\end{equation}
and the ANCOVA IV model with period-specific treatment-by-covariate interactions:
\begin{equation} \label{eq:int-model-tv}
	\begin{aligned}
		Y_{ijk} = \beta_{j} + \tau_{j}Z_{ij}+\widetilde{\bfX}_{ijk}\bfgamma_{j}
		+Z_{ij}\widetilde{\bfX}_{ijk}\bfeta_{j}+e_{ijk}.
	\end{aligned}
\end{equation}
Setting $\bfeta=\bm{0}$ and $\bfeta_{j}=\bm{0}$ respectively in the ANCOVA III and ANCOVA IV models, we obtain the ANCOVA I and ANCOVA II models as two special cases. For ANCOVA III and IV models, a reparameterization can facilitate the derivation of analytical results. Specifically, we separate the model components by treatment conditions and obtain the following reparameterized ANCOVA models:
\begin{equation*} 
	\begin{aligned}
		Y_{ijk} = (1-Z_{ij})\beta_{j} +(1-Z_{ij})\widetilde{\bfX}_{ijk}\bfgamma+ Z_{ij}\tau_{j}^*+Z_{ij}\widetilde{\bfX}_{ijk}\bfeta^* +e_{ijk},
	\end{aligned}
\end{equation*}
and
\begin{equation*} 
	\begin{aligned}
		Y_{ijk} = (1-Z_{ij})\beta_{j}+(1-Z_{ij})\widetilde{\bfX}_{ijk}\bfgamma_{j} + Z_{ij}\tau_{j}^*
		+Z_{ij}\widetilde{\bfX}_{ijk}\bfeta_{j}^*+e_{ijk},
	\end{aligned}
\end{equation*}
where $\tau_j^* = \beta_j+\tau_j$, $\bfeta^*=\bfgamma+\bfeta$, and $\bfeta_j^*=\bfgamma_j+\bfeta_j$. Our proposed ANCOVA estimators for $\tau_j$ are obtained via fitting either one of the above four working ANCOVA models using weighted least squares (WLS) with weights $w_{ijk}$ that are specified according to the estimand of interest. Beyond the weight specification, each of our estimators adopts a working independence assumption that does not explicitly account for the correlation between within-cluster observations during the estimation procedure. However, the correlation between observations will be addressed by the variance estimators in due course. For ease of reference, Table \ref{tab:estimators} provides a succinct summary of the different model formulations. 

\begin{table}[htbp] 
	\caption{Four model-assisted analyses of covariance estimators adjusting for baseline covariates in SW-CREs.}\label{tab:estimators}
	\vspace{-0.15in}
	\begin{center}
		\resizebox{\linewidth}{!}{
			\begin{tabular}{lccc} 
				\hline 
				\multirow{2}{*}{\textbf{Estimator}} & \multirow{2}{*}{\textbf{Mean model}} & \textbf{Period-specific} & \textbf{Treatment-by-} \\
				& & \textbf{covariate effects} & \textbf{covariate interactions}\\
				\hline
				ANCOVA I & $\beta_j+\tau_jZ_{ij}+(\bfX_{ijk}-\overline{\bfX}_j)\bfgamma$ & $\times$ & $\times$ \\
				\hline
				ANCOVA II & $\beta_j+\tau_jZ_{ij}+(\bfX_{ijk}-\overline{\bfX}_j)\bfgamma_j$ & $\checkmark$ & $\times$ \\ 
				\hline
				ANCOVA III & $\beta_j+\tau_jZ_{ij}+(\bfX_{ijk}-\overline{\bfX}_j)\bfgamma+Z_{ij}(\bfX_{ijk}-\overline{\bfX}_j)\bfeta$ & $\times$ & $\checkmark$ \\ 
				\hline
				ANCOVA IV & $\beta_j+\tau_jZ_{ij}+(\bfX_{ijk}-\overline{\bfX}_j)\bfgamma_j+Z_{ij}(\bfX_{ijk}-\overline{\bfX}_j)\bfeta_j$ & $\checkmark$ & $\checkmark$\\
				\hline
			\end{tabular}
		}
	\end{center}
\end{table}

Before we proceed, we first establish the following proposition, which indicates that the four ANCOVA estimators can be written in similar forms and, therefore, can be considered members of a larger family of estimators.

\begin{proposition} \label{eq:prop-1}
	Defining $w_j^1=\sumi Z_{ij}w_{ij}$ and $w_j^0=\sumi (1-Z_{ij})w_{ij}$ as the treatment-specific total weight per period, the WLS estimators for the regression coefficient (hence the weighted average treatment effect in period $j$) $\tau_j$, obtained from fitting ANCOVA I-IV, have the following closed-form expression:
	\begin{equation} \label{eq:estimator}
		\begin{aligned}
			\widehat{\Delta}_j=\frac{1}{w_j^1}\sum_{i:Z_{ij}=1}w_{ij}\overline{U}_{ij}(1)-\frac{1}{w_j^0}\sum_{i:Z_{ij}=0}w_{ij}\overline{U}_{ij}(0) =\overline{u}_j(1)-\overline{u}_j(0).
		\end{aligned}
	\end{equation}
	In particular, if we further define $\overline{y}_j(z)=\sum_{i:Z_{ij}=z} w_{ij}\overline{Y}_{ij}(z)/w_j^z$ and $\widetilde{\bfX}_j^z=\sum_{i:Z_{ij}=z} w_{ij}\widetilde{\bfX}_{ij}/w_j^z =\sum_{i:Z_{ij}=z}\sumk w_{ijk}\widetilde{\bfX}_{ijk}/w_j^z$ for $z\in\{0,1\}$. Then,
	\begin{enumerate}
		\item[(a)] for ANCOVA I, $\widehat{\Delta}_j=\widehat{\tau}_j$,  $\overline{U}_{ij}(z)=\overline{Y}_{ij}(z)-\widetilde{\bfX}_{ij}\widehat{\bfgamma}$, and $\overline{u}_j(z)=\overline{y}_j(z)-\widetilde{\bfX}_j^z\widehat{\bfgamma}$;
		\item[(b)] for ANCOVA II, $\widehat{\Delta}_j=\widehat{\tau}_j$, $\overline{U}_{ij}(z)=\overline{Y}_{ij}(z)-\widetilde{\bfX}_{ij}\widehat{\bfgamma}_j$, and $\overline{u}_j(z)=\overline{y}_j(z)-\widetilde{\bfX}_j^z\widehat{\bfgamma}_j$;
		\item[(c)] for ANCOVA III, $\widehat{\Delta}_j=\widehat{\tau}_j^*-\widehat{\beta}_j$, $\overline{U}_{ij}(1)=\overline{Y}_{ij}(1)-\widetilde{\bfX}_{ij}\widehat{\bfeta}^*$, $\overline{U}_{ij}(0)=\overline{Y}_{ij}(0)-\widetilde{\bfX}_{ij}\widehat{\bfgamma}$, $\overline{u}_j(1)=\overline{y}_j(1)-\widetilde{\bfX}_j^1\widehat{\bfeta}^*$, and $\overline{u}_j(0)=\overline{y}_j(0)-\widetilde{\bfX}_j^0\widehat{\bfgamma}$;
		\item[(d)] for ANCOVA IV, $\widehat{\Delta}_j=\widehat{\tau}_j^*-\widehat{\beta}_j$, $\overline{U}_{ij}(1)=\overline{Y}_{ij}(1)-\widetilde{\bfX}_{ij}\widehat{\bfeta}_j^*$, $\overline{U}_{ij}(0)=\overline{Y}_{ij}(0)-\widetilde{\bfX}_{ij}\widehat{\bfgamma}_j$,  $\overline{u}_j(1)=\overline{y}_j(1)-\widetilde{\bfX}_j^1\widehat{\bfeta}_j^*$, and $\overline{u}_j(0)=\overline{y}_j(0)-\widetilde{\bfX}_j^0\widehat{\bfgamma}_j$,
	\end{enumerate}
	where $\widehat{\tau}_j$, $\widehat{\tau}_j^*$, $\widehat{\beta}_j$, $\widehat{\bfgamma}$, $\widehat{\bfgamma}_j$, $\widehat{\bfeta}^*$, and $\widehat{\bfeta}_j^*$ are the estimated regression coefficients from fitting the respective working models.
\end{proposition}
Proposition \ref{eq:prop-1} suggests that all four ANCOVA estimators can be written in a similar form, which is the difference between weighted averages of cluster-period residualized potential outcomes. The nonparametric estimator, where no covariate adjustments are considered, is also a special member of \eqref{eq:estimator} with $\overline{U}_{ij}(z)=\overline{Y}_{ij}(z)$ and $\overline{u}_j(z)=\overline{y}_j(z)$. The proof of Proposition \ref{eq:prop-1} is given in Web Appendix A. With Proposition \ref{eq:prop-1}, we can straightforwardly show that, under Assumptions \ref{asp:sutva}--\ref{asp:rand}, each ANCOVA estimator of $\tau$, defined as a weighted average of $\widehat{\Delta}_j$, $j=1,\ldots,J$, i.e.,  
\begin{align} \label{eq:total-estimator}
	\widehat{\tau} = \frac{\sumj w_j\widehat{\Delta}_j}{\sumj w_j}.
\end{align}
is consistent with the target estimand $\tau$ even if the working ANCOVA models are incorrectly specified. In \eqref{eq:total-estimator}, $\widehat{\Delta}_j$ is equal to $\widehat{\tau}_j$ when ANCOVA I and II are considered, and $\widehat{\tau}_j^*-\widehat{\beta}_j$ when the reparameterized ANCOVA III and IV are considered, according to Proposition \ref{eq:prop-1}.

\subsection{Theoretical properties}\label{sec:theory}

We study the theoretical properties of the ANCOVA estimators for estimating the general class of estimands, $\tau$. The technical presentation below serves two purposes. First, it emphasizes the key role of staggered randomization in determining the large-sample distribution of $\widehat{\tau}$, a unique complexity arising from the stepped wedge design compared to simpler parallel-arm designs. Second, the technical development directly motivates a design-based variance estimator for $\widehat{\tau}$ to quantify its uncertainty and facilitate the construction of confidence intervals and hypothesis tests. Under the staggered rollout randomization scheme, we formalize the theoretical properties of $\widehat{\tau}$ under the asymptotic regime similar to that in \citet{Middleton2015}, \citet{Li2017} and \citet{Schochet2021}, where an increasing sequence of finite populations are considered with the number of clusters, $I\rightarrow\infty$. In our case, the number of rollout periods, $J$, is considered fixed. Furthermore, we assume that the number of clusters randomized to the treatment condition in period $j$ increases proportionally, that is, $I_j/I = e_j$ as $I\rightarrow\infty$. Finally, we assume that the number of individuals in each cluster and the corresponding weight remains relatively balanced for each period (no clusters with a dominating number of individuals or weights in any rollout period), and neither varies as a function of $I$. The last assumption is to prevent ill-mannered behaviors of the ANCOVA estimators. 

For stepped wedge designs, an important consideration is that treatment assignments for a cluster and between any pairs of clusters across rollout periods are correlated, which differs from a parallel-arm or blocked cluster randomization scheme (where randomization is conducted independently within each block). Therefore, these correlations in the treatment assignment must be addressed to present a finite population Central Limit Theorem (CLT) for $\widehat{\tau}$. Similar to \citet{Athey2022} and \citet{Roth2021}, we introduce an alternative perspective of viewing different treatment adoption times as different treatment arms and potential outcomes for a cluster across $J$ rollout periods as a $J$-dimensional potential outcome vector, which leverages the random assignment to multiple arms uniquely determined by the treatment adoption time. This is especially important because we are interested in obtaining the appropriate randomization distribution of the estimator $\widehat{\tau}$. 

Specifically, we let $A_i=a$, $a\in\calA=\{1,\ldots,J,J+1\}$, denote the adoption date of the treatment for cluster $i$, and we have $G_i^a=\bbI\{A_i=a\}$, which is equal to one if cluster $i$ adopts the treatment at period $a$. Note that cluster $i$ does not receive treatment throughout the rollout phase if $A_i=J+1$. Under this perspective, the $J$-dimensional weighted average for cluster $i$ is $\overline{\bfY}_i^a=(\overline{Y}_{i1}^a,\ldots,\overline{Y}_{iJ}^a)'$, where $\overline{Y}_{ij}^a = \sumk w_{ijk}Y_{ijk}^a/\sumk w_{ijk}$. We further define $\overline{\bfY}^a=(\overline{Y}_1^a,\ldots,\overline{Y}_J^a)'$, where $\overline{Y}_j^a=\sumi w_{ij}\overline{Y}_{ij}^a/\sumi w_{ij}$. Importantly, while the randomization distribution of the observed quantities is dictated by the randomness of the adoption date and hence $A_i$, our causal estimands are defined based on $\overline{Y}_j(1)$ and $\overline{Y}_j(0)$ due to Assumption \ref{asp:dur}. Therefore, bridging the connection between these two versions of the potential outcomes is important. To this end, we write $\overline{Y}_j(1)$ and $\overline{Y}_j(0)$ as a function of $\overline{Y}_j^a$ to highlight the implicit role of the adoption date in the target estimands, to motivate an explicit reformulation of each ANCOVA estimator based on the adoption date, and eventually to facilitate the derivation of the randomization distribution of each ANCOVA estimator. Under Assumption \ref{asp:dur}, we can re-express components of the weighted average treatment effect of period $j$ as
\begin{align*}
	\overline{Y}_j(1)&=\frac{\suma\bbI\{a\leq j\}w_j\overline{Y}_j^a}{\suma\bbI\{a\leq j\}w_j}=\frac{\suma\bbI\{a\leq j\}\overline{Y}_j^a}{\suma\bbI\{a\leq j\}},\\
	\overline{Y}_j(0)&=\frac{\suma\bbI\{a>j\}w_j\overline{Y}_j^a}{\suma\bbI\{a>j\}w_j}=\frac{\suma\bbI\{a>j\}\overline{Y}_j^a}{\suma\bbI\{a>j\}},
\end{align*}
which leads to the reformulation of $\tau$ as a linear combination of $\overline{Y}_j^a$'s, i.e.,
\begin{align} \label{eq:estimand-reexp}
	\tau = \sumj\suma {B_j^a}\overline{Y}_j^a,
\end{align}
where
\begin{align*}
	B_j^a=\frac{w_j}{\sumj w_j}\left\{\frac{\bbI\{a\leq j\}}{\suma\bbI\{a\leq j\}}-\frac{\bbI\{a> j\}}{\suma\bbI\{a> j\}}\right\}.
\end{align*}
Similarly, the estimator, $\widehat{\tau}$, can also be re-expressed in terms of random components under the perspective of multiple-arm assignment. Analogous to the definition of $\overline{Y}_{ij}^a$ and $\overline{Y}_j^a$, we define $\overline{U}_j^a=\sumi w_{ij}\overline{U}_{ij}^a/\allowbreak\sumi w_{ij}$, where 
\begin{enumerate}
	\item[(a)] for ANCOVA I, $\overline{U}_{ij}^a=\overline{Y}_{ij}^a-\widetilde{\bfX}_{ij}\widehat{\bfgamma}$;
	\item[(b)] for ANCOVA II, $\overline{U}_{ij}^a=\overline{Y}_{ij}^a-\widetilde{\bfX}_{ij}\widehat{\bfgamma}_j$;
	\item[(c)] for ANCOVA III, $\overline{U}_{ij}^a=\overline{Y}_{ij}^a-\widetilde{\bfX}_{ij}(\bbI\{a\leq j\}\widehat{\bfeta}^*+\bbI\{a> j\}\widehat{\bfgamma})$;
	\item[(d)] for ANCOVA IV, $\overline{U}_{ij}^a=\overline{Y}_{ij}^a-\widetilde{\bfX}_{ij}(\bbI\{a\leq j\}\widehat{\bfeta}_j^*+\bbI\{a> j\}\widehat{\bfgamma}_j)$.
\end{enumerate}
If we further write $\overline{u}_j^a=\sumi w_{ij}G_i^a\overline{U}_{ij}^a/\sumi w_{ij}G_i^a$, then we have the following re-expressions for random components:
\begin{align*}
	\overline{u}_j(1)=\frac{\suma\bbI\{a\leq j\}\left(\sumi w_{ij}G_i^a\right)\overline{u}_j^a}{\suma\bbI\{a\leq j\}\left(\sumi w_{ij}G_i^a\right)},~~
	\overline{u}_j(0)=\frac{\suma\bbI\{a>j\}\left(\sumi w_{ij}G_i^a\right)\overline{u}_j^a}{\suma\bbI\{a>j\}\left(\sumi w_{ij}G_i^a\right)}.
\end{align*}
Hence $\widehat{\tau}$ can be re-expressed as
\begin{align} \label{eq:estimator-reexp}
	\widehat{\tau} = \sumj\suma b_j^a\overline{u}_j^a,
\end{align}
where
\begin{align*}
	b_j^a=\frac{w_j}{\sumj w_j}\left\{\frac{\bbI\{a\leq j\}\left(\sumi w_{ij}G_i^a\right)}{\sum_a\bbI\{a\leq j\}\left(\sumi w_{ij}G_i^a\right)}-\frac{\bbI\{a> j\}\left(\sumi w_{ij}G_i^a\right)}{\sum_a\bbI\{a> j\}\left(\sumi w_{ij}G_i^a\right)}\right\}.
\end{align*}
We proceed to first define the intermediate quantities, $\widetilde{U}_{ij}^a$, by assuming that estimates of associated parameter vectors, $\widehat{\bfgamma}$, $\widehat{\bfgamma}_j$, $\widehat{\bfeta}^*$, and $\widehat{\bfeta}_j^*$, are substituted by their corresponding known values, $\bfgamma$, $\bfgamma_j$, $\bfeta^*$, and $\bfeta_j^*$, which are WLS coefficient vectors that would be obtained if the full set of potential outcomes is available. Let $I^a=I_a-I_{a-1}$ denote the number of clusters with the adoption date $a$. We then define vectors of period-level random components, $\overline{\bft}^a=(w\widetilde{u}_1^a,\ldots,w\widetilde{u}_J^a,\overline{w}_1^a,\ldots,\overline{w}_J^a)'$, where $w\widetilde{u}_j^a=(I^a)^{-1}\sumi G_i^aw_{ij}\widetilde{U}_{ij}^a$ and $\overline{w}_j^a=(I^a)^{-1}\sumi\allowbreak w_{ij}G_i^a$, and vectors of period-level average components, $\overline{\bfT}^a=(w\widetilde{U}_1^a,\ldots,w\widetilde{U}_J^a,\overline{w}_1,\ldots,\overline{w}_J)'$, where $w\widetilde{U}_j^a=I^{-1}\sumi w_{ij}\widetilde{U}_{ij}^a$ and $\overline{w}_j=I^{-1}\sumi w_{ij}$, with cluster-level vectors, $\overline{\bfT}_i^a=(w_{i1}\widetilde{U}_{i1}^a,\ldots,\allowbreak w_{iJ}\widetilde{U}_{iJ}^a,\allowbreak w_{i1},\ldots,w_{iJ})'$ such that $\overline{\bfT}^a=I^{-1}\sumi\overline{\bfT}_i^a$. Also, we define covariance matrices, $\bfS_T^a=(I-1)^{-1}\sumi(\overline{\bfT}_i^a-\overline{\bfT}^a)(\overline{\bfT}_i^a-\overline{\bfT}^a)'$, as well as cross-product matrices, $\bfS_T^{a,a'}=(I-1)^{-1}\sumi(\overline{\bfT}_i^a-\overline{\bfT}^a)(\overline{\bfT}_i^{a'}-\overline{\bfT}^{a'})'$. Similar to $\widetilde{U}_{ij}^a$, we define the intermediate quantities $\widetilde{U}_{ij}(z)$ by replacing estimates of associated parameter vectors with corresponding know values. From the re-expression of $\widehat{\tau}$ in \eqref{eq:estimator-reexp}, we can obtain cluster-level intermediate vectors
\begin{align*}
	\widetilde{\bfU}_i^a=\left(\widetilde{\bfU}_i(1)-\widetilde{\bfU}(1)\right)\left(\otimes_{j=1}^J\bbI\{a\leq j\}\right)+\left(\widetilde{\bfU}_i(0)-\widetilde{\bfU}(0)\right)\left(\otimes_{j=1}^J\bbI\{a> j\}\right),
\end{align*}
where $\widetilde{\bfU}_i^a=(\widetilde{U}_{i1}^a,\ldots,\widetilde{U}_{iJ}^a)'$, $\widetilde{\bfU}_i(z)=(\widetilde{U}_{i1}(z),\ldots,\allowbreak\widetilde{U}_{iJ}(z))'$, and $\widetilde{\bfU}(z)=(\widetilde{U}_1(z),\ldots,\widetilde{U}_J(z))'$ with $\widetilde{U}_j(z)=\sumi w_{ij}\allowbreak\widetilde{U}_{ij}(z)/\sumi w_{ij}$; `$\otimes$' is the block diagonal operator, and therefore $\otimes_{j=1}^J\bbI\{a\leq j\}$ and $\otimes_{j=1}^J\bbI\{a> j\}$ are $J\times J$ diagonal matrices, with the $j$-th diagonal element being $\bbI\{a\leq j\}$ and $\bbI\{a> j\}$, respectively. In addition, we have $J\times J$ cluster-level adjusted diagonal weight matrices
\begin{align*}
	\widetilde{\bfW}_i^a=\otimes_{j=1}^J\left(\frac{I^aw_{ij}}{I_j\overline{w}_j}\bbI\{a\leq j\}-\frac{I^aw_{ij}}{(I-I_j)\overline{w}_j}\bbI\{a>j\}\right).
\end{align*}
Finally, we have covariance matrices, $\bfS_{\widetilde{U},\widetilde{W}}^a=(I-1)^{-1}\sumi\widetilde{\bfU}_i^a\widetilde{\bfW}_i^a\widetilde{\bfW}_i^a\widetilde{\bfU}_i^a{}'$, cross-product matrices, $\bfS_{\widetilde{U},\widetilde{W}}^{a,a'}=(I-1)^{-1}\sumi\widetilde{\bfU}_i^a\widetilde{\bfW}_i^a\widetilde{\bfW}_i^{a'}\widetilde{\bfU}_i^{a'}{}'$, weighted sample covariance matrices for covariates, $\bfS_{\bfX,j}={I}^{-1}\sumi\sumk w_{ijk}\widetilde{\bfX}_{ijk}'\widetilde{\bfX}_{ijk}$, and weighted cross-product vectors for the covariates and potential outcomes, $\bfS_{\bfX,Y,j}(z)={I}^{-1}\sumi\sumk w_{ijk}\widetilde{\bfX}_{ijk}'Y_{ijk}(z)$. Based on these intermediate quantities, we now present the finite population CLT for $\widehat{\tau}$, and the proof is provided in Web Appendix B.
\begin{theorem} \label{thm:1}
	Under Assumptions \ref{asp:sutva} - \ref{asp:rand}, and further assuming the following regularity conditions for $a\in\calA$ and $j\in\{1,\ldots,J\}$:
	\begin{itemize}
		\item[(i)] Define $m_j^a(\widetilde{U})=\max_{1\leq i\leq I}\left\{w_{ij}(\widetilde{U}_{ij}^a-\widetilde{U}_j^a)\right\}^2$, $v_j^a(\widetilde{U})=(I-1)^{-1}\sumi\left\{w_{ij}(\widetilde{U}_{ij}^a-\widetilde{U}_j^a)\right\}^2$, and as $I\rightarrow\infty$,
		\begin{align*}
			\max_{a\in\calA}\max_{1\leq j\leq J}\frac{m_j^a(\widetilde{U})}{I^av_j^a(\widetilde{U})} \rightarrow 0.
		\end{align*}
		\item[(ii)] Define $m_j(w)=\max_{1\leq i\leq I}\left(w_{ij}-\overline{w}_j\right)^2$, $v_j(w)=(I-1)^{-1}\sumi\left(w_{ij}-\overline{w}_j\right)^2$, and as $I\rightarrow\infty$,
		\begin{align*}
			\max_{a\in\calA}\max_{1\leq j\leq J}\frac{m_j(w)}{I^av_j(w)} \rightarrow 0.
		\end{align*}
		\item[(iii)] Define 
		\begin{align*}
			m_{jl}(\widetilde{\bfX})=\max_{1\leq i\leq I}\left(\frac{w_{ij}}{\overline{w}_j}[\widetilde{\bfX}_{ij}]_l\right)^2,~~\text{and}~~
			v_{jl}(\widetilde{\bfX})=\frac{1}{I-1}\sumi\frac{w_{ij}^2}{\overline{w}_j^2}[\widetilde{\bfX}_{ij}]_l^2,
		\end{align*}
		for $l\in\{1,\ldots,p\}$, and as $I\rightarrow\infty$,
		\begin{align*}
			\max_{a\in\calA}\max_{1\leq j\leq J}\frac{m_{jl}(\widetilde{\bfX})}{I^av_{jl}(\widetilde{\bfX})} \rightarrow 0.
		\end{align*}
		\item[(iv)] Assume $\bfS_T^a$, $\bfS_T^{a,a'}$, $\bfS_{\widetilde{U},\widetilde{W}}^a$, $\bfS_{\widetilde{U},\widetilde{W}}^{a,a'}$, and $\bfS_{\bfX,j}$ have finite, and positive definite limiting values; the correlation matrix of $\overline{\bft}^a$ has a finite limiting value $\bfR_T$, defined in Web Appendix B.
		\item[(v)] Assume $\sumi w_{ij}\widetilde{U}_{ij}(1)\neq 0$ or $\sumi w_{ij}\widetilde{U}_{ij}(0)\neq 0$, for some $j$.
	\end{itemize}
	Then, under any specification of the weight $w_{ijk}$ that meet the above regularity conditions, as $I\rightarrow\infty$, $\widehat{\tau}$ is a consistent estimator for $\tau$ and 
	\begin{align*}
		\frac{\widehat{\tau}-\tau}{\sqrt{\mathrm{var}(\widehat{\tau})}}\xrightarrow{d}\calN\left(0,1\right),
	\end{align*}
	where $\mathrm{var}(\widehat{\tau}) = \bfvarpi'\bfSigma_{\tau}\bfvarpi$ with
	\begin{align} \label{eq:estimator-cov-mat}
		\bfSigma_{\tau} 
		=&\suma\frac{1}{I^a}\bfS_{\widetilde{U},\widetilde{W}}^a-\sum_{a,a'\in\calA}\frac{1}{I}\bfS_{\widetilde{U},\widetilde{W}}^{a,a'},
	\end{align}
	and $\bfvarpi=(\varpi_1,\ldots,\varpi_J)'$, $\varpi_j=w_j/\sumj w_j$.
\end{theorem}

To give a more concrete context of the technical development in Theorem \ref{thm:1}, we provide the interpretation of each regularity condition. Conditions (i)-(v) are adapted from conditions required by Theorems 1 and 2 in \citet{Schochet2021} for blocked cluster randomized designs. By definition, conditions (i)-(iii) are Lindeberg-type conditions for controlling the tail behaviors, which is required for invoking the finite population CLT in Theorem 4 of \citet{Li2017}. They regulate the ratio between the maximum squared distance and the finite population variance of a population characterized by their treatment adoption time $a$. Particularly, condition (i) is defined for the residualized potential outcomes, condition (ii) is defined for the cluster-period weights, whereas condition (iii) is defined for the collection of covariates adjusted for in the ANCOVA working models. Condition (iv) ensures that limiting values of covariance matrices of residualized potential outcomes, sampling weights, and covariates exist. Condition (v) is required to guarantee that the covariance matrix of $\bfh(\overline{\bft}^a)$ obtained from applying the delta method for finite population inference \citep{Pashley2022} is positive definite, where $\bfh(\cdot)$ is a function of $\overline{\bft}^a$ given in Web Appendix B. The proof of Theorem \ref{thm:1} starts with introducing intermediate quantities by assuming the associated parameter vectors are known and establishing a CLT with known parameters. These known parameters are WLS coefficient vectors that would be obtained if the full set of potential outcomes were available. We then proceed to show that estimates of these vectors converge to the same asymptotic value as the known parameters, which subsequently leads to the result that $\widehat{\tau}$ converges to a standard normal distribution. Finally, although Theorem \ref{thm:1} is general with respect to the form and (finite) dimension of the covariates, it also applies to the nonparametric estimator without covariate adjustment where the covariate regression parameters are uniformly set to zero.

\begin{remark}
	Of note, the first term of \eqref{eq:estimator-cov-mat} is a summation of separate covariance matrices of cluster-level average model residuals for different groups characterized by the respective treatment adoption times, and it is the counterpart of the variance developed under the conventional two-arm randomized experiment with a univariate outcome \citep[Chapter 6]{Imbens2015}. We notice that the second term can be rearranged as
	\begin{align*}
		\sum_{a,a'\in\calA}\frac{1}{I}\bfS_{\widetilde{U},\widetilde{W}}^{a,a'}=\frac{1}{I(I-1)}\sumi\left(\suma\widetilde{\bfU}_i^a\widetilde{\bfW}_i^a\right)\left(\sum_{a'\in\calA}\widetilde{\bfU}_i^{a'}\widetilde{\bfW}_i^{a'}\right)',
	\end{align*}
	which is the covariance matrix of cluster-level average residualized potential outcomes among different treatment adoption time groups. A closer look into the diagonal entries of \eqref{eq:estimator-cov-mat} with the above rearrangement reveals some interesting connections between the stepped wedge design and the blocked cluster randomized design in \citet{Schochet2021}. Specifically, we first study the weight matrix $\widetilde{\bfW}_i^a$, where if $a\leq j$, then $I^a/I_j$ is the ratio between the number of clusters starting to adopt the treatment to the total number of treated clusters in period $j$; if $a>j$, then $I^a/(I-I_j)$ is the ratio between the number of clusters scheduled to adopt the treatment in period $a$ to the total number of untreated clusters in period $j$. In a blocked cluster randomized design with a balanced number of clusters in each block, where clusters in different blocks are independently randomized to the treatment or control arms, the number of treatment adoption times can only be now ($a=0$) or never ($a=\infty$) ($\calA=\{0,\infty\}$). The immediate implication under this setup is that $I^a=I_j$ when $a=0$, and $I^a=I-I_j$ when $a=\infty$. Therefore, the $j$-th diagonal entry from the first term of \eqref{eq:estimator-cov-mat} becomes
	\begin{align*}
		\frac{1}{I_j}\frac{1}{I-1}\sumi\frac{w_{ij}^2}{\overline{w}_j^2}\left\{\widetilde{U}_{ij}(1)-\widetilde{U}_j(1)\right\}^2+\frac{1}{I-I_j}\frac{1}{I-1}\sumi\frac{w_{ij}^2}{\overline{w}_j^2}\left\{\widetilde{U}_{ij}(0)-\widetilde{U}_j(0)\right\}^2,
	\end{align*}
	since $\widetilde{U}_{ij}^0=\widetilde{U}_{ij}(1)$ and $\widetilde{U}_{ij}^\infty=\widetilde{U}_{ij}(0)$. Under the same blocked cluster randomized design, the $j$-th diagonal entry of the second term of \eqref{eq:estimator-cov-mat} after the above rearrangement becomes
	\begin{align*}
		\frac{1}{I(I-1)}\sumi \frac{w_{ij}^2}{\overline{w}_j^2}\left[\widetilde{U}_{ij}(1)-\widetilde{U}_{ij}(0)-\left\{\widetilde{U}_j(1)-\widetilde{U}_j(0)\right\}\right]^2,
	\end{align*}
	which is precisely the variance expression of the average treatment effect estimator across clusters in block $j$. Putting the first and second terms together, we obtain the finite population variance of $\widehat{\tau}_j$ given in \citet{Schochet2021}. From this standpoint, our variance in Theorem \ref{thm:1} generalizes the variance developed for blocked cluster randomized experiments.
\end{remark}

\subsection{Variance estimation} 

The estimation of $\mathrm{var}(\widehat{\tau})$ hinges upon that of $\bfSigma_{\tau}$, the covariance matrix of $(\widehat{\tau}_1,\ldots,\widehat{\tau}_J)'$, since the normalized weight vector, $\bfvarpi$, is known. The second term of $\bfSigma_{\tau}$, as previously mentioned, is the variance of the average treatment effect estimators across clusters in period $j$, which is generally inestimable because each cluster can only be randomized to a specific treatment adoption time in practice. The first term, the summation of separate covariance matrices of cluster-level average model residuals for different treatment adoption time groups, can be estimated via a consistent design-based (DB) plug-in estimator. Specifically, the covariance matrix component for the group with treatment adoption time $a$, $\bfS_{\widetilde{U},\widetilde{W}}^a$ can be estimated by
\begin{align} \label{eq:db-var}
	\widehat{\bfS}_{U,W}^a=\frac{1}{I^a-1}\sum_{i:A_i=a}\widehat{\bfU}_i^a\widehat{\bfW}_i^a\widehat{\bfW}_i^a\widehat{\bfU}_i^a{}',
\end{align}
where $\widehat{\bfU}_i^a=(\widehat{U}_{i1}^a,\ldots,\widehat{U}_{iJ}^a)'$, and $\widehat{U}_{ij}^a=(\overline{U}_{ij}(1)-\overline{u}_j(1))(\otimes_{j=1}^J\bbI\{a\leq j\})+(\overline{U}_{ij}(0)-\overline{u}_j(0))\allowbreak\times(\otimes_{j=1}^J\bbI\{a> j\})$. 
The estimator for the weight matrix is 
\begin{align*}
	\widehat{\bfW}_i^a=\otimes_{j=1}^J\left(\frac{I^aw_{ij}}{I_j\overline{w}_j^1}\bbI\{a\leq j\}-\frac{I^aw_{ij}}{(I-I_j)\overline{w}_j^0}\bbI\{a>j\}\right),
\end{align*}
where $\overline{w}_j^1=I_j^{-1} w_j^1$ and $\overline{w}_j^0=(I-I_j)^{-1}w_j^0$. The DB estimator for $\mathrm{var}(\widehat{\tau})$ thus is
\begin{align} \label{eq:var-est-db}
	\widehat{\mathrm{var}}_{DB}(\widehat{\tau})=\bfvarpi'\left(\suma\frac{1}{I^a}\widehat{\bfS}_{U,W}^a\right)\bfvarpi.
\end{align}
Note that since the second term,
\begin{align*}
	\bfvarpi'\left(\sum_{a,a'\in\calA}\frac{1}{I}\bfS_{\widetilde{U},\widetilde{W}}^{a,a'}\right)\bfvarpi=\frac{1}{I(I-1)}\sumi\left(\bfvarpi'\suma\widetilde{\bfU}_i^a\widetilde{\bfW}_i^a\right)^2\geq 0,
\end{align*}
and is generally unestimable, the DB estimator is expected to be conservative. 
\begin{remark}
	Due to the construction of the design-based variance estimator, equation \eqref{eq:db-var} is undefined when a treatment sequence only contains a single cluster or $I_a=1$ for some $a\in\calA$. To ensure a feasible design-based variance estimator in this special case, the expression in \eqref{eq:db-var} can be modified to $\widehat{\bfS}_{U,W}^a=(I^a)^{-1}\sum_{i:A_i=a}\widehat{\bfU}_i^a\widehat{\bfW}_i^a\widehat{\bfW}_i^a\widehat{\bfU}_i^a{}'$. The resulting modified variance estimator remains conservative and asymptotically unbiased as $I\rightarrow\infty$.
\end{remark}

The variance of $\widehat{\tau}$ can also be estimated using the cluster-robust standard errors (CRSE) \citep{Liang1986}, which assumes the working independence between clusters but allows errors within the same cluster to be arbitrarily correlated. Slightly different from the DB variance estimator, which is directly motivated by the randomization-based variance expression under a finite population framework, the CRSE estimator was originally developed for generalized estimating equations estimators under a super population sampling-based framework \citep{ding2017bridging}, assuming the clusters are sampled from an infinite population of clusters. While the asymptotic property of the CRSE estimator may be challenging to study under a finite population perspective in the context of the staggered cluster randomization scheme, we empirically compare the CRSE and the DB standard error estimators in our simulations, as the former is often recognized as a potentially more accessible strategy (due to software availability for generalized estimating equations) to quantify the uncertainty of $\widehat{\tau}$. We will also return to a more detailed discussion about these different variance estimators for SW-CREs in Section \ref{sec:discussion}.

Using individual-level data, the CRSE estimator for the covariance matrix of WLS ANCOVA model parameters is:
\begin{align*}
	\bfE=\left(\sum_{i=1}^I\bfD_i'\bfW_i\bfD_i\right)^{-1}\left(\sumi\bfD_i'\bfW_i\widehat{\bfe}_i\widehat{\bfe}_i'\bfW_i\bfD_i\right)\left(\sum_{i=1}^I\bfD_i'\bfW_i\bfD_i\right)^{-1},
\end{align*}
where $\bfD_i$ is the design matrix including all individuals in cluster $i$ specified according to the ANCOVA model adopted, with each row corresponding to an individual observation, $\bfW_i$ is the diagonal weight matrix containing all individuals in cluster $i$, and $\widehat{\bfe}_i$ is the vector of estimated WLS residuals in cluster $i$. Notice that this is essentially the robust sandwich variance estimator for weighted generalized estimating equations under the independent working correlation structure; a more general form of this variance can be found, for example, in \citet{preisser2002performance}. Here, we use ANCOVA I as an example to explain that the CRSE for $\widehat{\tau}$ could be obtained, and the other three models follow similarly. In particular, under ANCOVA I, $\bfE$ is a $(2J+p)\times(2J+p)$ symmetric matrix, with
\begin{align*}
	\bfE=\left(\begin{array}{ccc}
		\bfE_{11} & \bfE_{12} & \bfE_{13} \\
		\bfE_{21} & \bfE_{22} & \bfE_{23} \\
		\bfE_{31} & \bfE_{32} & \bfE_{33} 
	\end{array}\right),
\end{align*}
and $\bfE_{22}$ is the CRSE estimator for $\bfSigma_{\tau}$. Thus, the CRSE estimator for $\mathrm{var}(\widehat{\tau})$ is $\widehat{\mathrm{var}}_{CRSE}(\widehat{\tau})=\bfvarpi'\bfE_{22}\bfvarpi$. We proceed to compare the DB and CRSE estimators via simulations to inform their applications to SW-CREs.

\section{Simulation Studies} \label{sec:simulation}

\subsection{Simulation study I}

We conduct simulation studies to evaluate the finite-sample operating characteristics of the ANCOVA estimators and compare across different model formulations (ANCOVA I-IV), as well as the unadjusted estimators, where no covariate effects are considered (ANCOVA I with $\bfgamma=\bm{0}$). In simulation study I, we consider a fixed total number of $J=5$ rollout periods, with one pre-rollout ($j=0$) and one post-rollout period ($j=J+1$), and three settings featuring different numbers of clusters with $I\in\{18,30,60,120\}$. In each rollout period, $I/(J+1)$ previously untreated clusters are randomized to the treatment arm, and the remaining untreated clusters after the rollout period $J$ will be assigned to the treatment arm in the post-rollout period. For $j\in\{0,1,\ldots,J+1$\}, we simulate the cluster-period size from $N_{ij}\sim\calU(10,90)+2.5(j+1)^2$ rounded to the nearest integer, and $\calU$ stands for the uniform distribution; $X_{ij1}\sim \calB(0.5)$ represents an exogenous cluster-period-level summary variable following the Bernoulli distribution with $\bbP(X_{ij1}=1)=0.5$, and $X_{ijk2}\sim i/I+\calU(-1,1)$ is an individual-level covariate. The potential outcomes are generated as
\begin{align*}
	Y_{ijk}(0) &=\frac{j+1}{J+2}+X_{ij1}+\left(X_{ijk2}-\overline{X}_{j2}\right)^2 + c_i +e_{ijk}, \displaybreak[0]\\
	Y_{ijk}(1) &= Y_{ijk}(0)+\frac{2N_{ij}I}{(J+2)^{-1}\sum_{j=0}^{J+1} N_j}+0.5X_{ij1}+\left(X_{ijk2}-\overline{X}_{j2}\right)^3,
\end{align*}
where $\overline{X}_{j2}$ is the weighted average of $X_{ijk2}$ in period $j$ depending on weights specified for each estimand, $c_i$ is the cluster-specific random effect of cluster $i$, and $e_{ijk}$ is the individual-specific error independent from $c_i$ and other components, with $c_i\sim\calN(0,\sigma_c^2)$ and $e_{ijk}\sim\calN(0,\sigma_e^2)$. Here, we have $\sigma_c^2=0.1$ and $\sigma_e^2=0.9$, leading to a common intracluster correlation coefficient (ICC) of $\sigma_c^2/(\sigma_c^2+\sigma_e^2)=0.1$. The generating process of potential outcomes is nonlinear in covariates, as it is designed to examine the model-robustness of proposed estimators. We then fit models \eqref{eq:gen-model} - \eqref{eq:int-model-tv} with covariate vector $\widetilde{\bfX}_{ijk}$ as the centered version of $\bfX_{ijk}=(X_{ij1},X_{ijk2})$. We further assess the impact of including the cluster-period size $N_{ij}$ as an additional covariate and discuss those results later. 

For each scenario, we simulate 1,000 stepped wedge cluster-randomized experiments and evaluate finite-sample properties of proposed estimators by calculating their relative bias (BIAS), root mean square error (RMSE), and empirical coverage percentages of 95\% confidence intervals (CIs) using the DB and CRSE estimators. Additionally, we compare the average standard errors (ASEs) from the DB and CRSE approaches with the corresponding Monte Carlo or empirical standard errors (ESEs) of proposed estimators.

Simulation results for estimating $\tau^{ind}$ and $\tau^{cell}$ are presented in Tables \ref{tab:sim-1-ate-1} and \ref{tab:sim-1-ate-3}, and results of estimating $\tau^{period}$ are available in Web Table 1 of Web Appendix C. Due to informative cluster-period size, the true estimands differ such that $\tau^{ind}=0.729$, $\tau^{period}=0.638$, and $\tau^{cell}=0.525$. For each estimand, as the number of clusters, $I$, increases, the relative bias and RMSE decrease for all five compared estimators, corroborating the consistency property of proposed approaches (also shown in Web Figure 1). A general observation is that, with sufficient clusters (e.g. $I=120$), both the DB and CRSE methods provide standard error estimates generally closer to the ESE. However, properties of variance estimation approaches can differ substantially with a relatively limited number of clusters. For the DB approach, we found that it is more conservative for the unadjusted, ANCOVA I and II estimators, and the conservativeness is more pronounced when the number of clusters is relatively small ($I\in\{18,30\}$). For ANCOVA III, which includes treatment-by-covariate interactions but without period-specific covariate effects, the DB approach yields ASEs close to ESEs, with coverage percentages of 95\% confidence intervals close to the nominal level, regardless of the number of clusters. For ANCOVA IV, which includes period-specific interactions, the DB approach slightly underestimates the variance when $I$ is relatively small but yields relatively accurate results when $I$ increases ($I\in\{60,120\}$). With a limited number of clusters, underestimating the variance from the DB approach for ANCOVA IV is unsurprising because the working model includes an increasing number of parameters compared to other candidate estimators.

\begin{table}[htbp] 
	\caption{Results for the individual-average treatment effect ($\tau^{ind}$) from simulation study I comparing performances of five estimators, UN = unadjusted, AN I-IV = ANCOVA I-IV. Evaluation metrics: BIAS = relative bias; RMSE = root mean squared error; ESE = empirical standard error; ASE = average standard error, where, DB = standard errors via the design-based plug-in estimator, and CRSE = cluster-robust standard errors; Coverage = empirical coverage of 95\% confidence intervals over 1,000 simulation replications.}\label{tab:sim-1-ate-1}
	\par
	\begin{center}
			\begin{tabular}{cc r cc cc cc} 
				\hline 
				& & & & & \multicolumn{2}{c}{ASE} & \multicolumn{2}{c}{Coverage} \\
				\cline{6-9} \\[-1.5ex]
				$I$  & Estimator  & \multicolumn{1}{c}{BIAS} & RMSE & ESE & DB & CRSE & DB & CRSE \\
				\hline\\[-2ex]
				18  & UN  & -0.022 & 0.159 & 0.180 & 0.213 & 0.174 & 0.955 & 0.913 \\
				&  AN I  & -0.013 & 0.101 & 0.131 & 0.163 & 0.134 & 0.977 & 0.941 \\
				&  AN II  & -0.013 & 0.116 & 0.143 & 0.160 & 0.133 & 0.962 & 0.909 \\
				&  AN III  & -0.010 & 0.097 & 0.129 & 0.133 & 0.110 & 0.942 & 0.892 \\
				& AN IV  & -0.003 & 0.107 & 0.137 & 0.112 & 0.090 & 0.869 & 0.772 \\
				\hline\\[-2ex]
				30  & UN  & -0.019 & 0.129 & 0.145 & 0.160 & 0.143 & 0.958 & 0.932 \\
				&  AN I  & -0.009 & 0.083 & 0.108 & 0.124 & 0.112 & 0.973 & 0.950 \\
				&  AN II  & -0.012 & 0.094 & 0.118 & 0.125 & 0.113 & 0.955 & 0.923 \\
				&  AN III  & -0.006 & 0.079 & 0.105 & 0.103 & 0.093 & 0.934 & 0.900 \\
				&  AN IV  & -0.006 & 0.085 & 0.110 & 0.093 & 0.081 & 0.880 & 0.834 \\
				\hline\\[-2ex]
				60  & UN 	& -0.005 & 0.085 & 0.095 & 0.113 & 0.107 & 0.975 & 0.968 \\
				&  AN I & -0.001 & 0.054 & 0.070 & 0.089 & 0.084 & 0.982 & 0.975 \\
				&  AN II  & -0.001 & 0.059 & 0.074 & 0.091 & 0.087 & 0.983 & 0.975 \\
				&  AN III  & 0.001 & 0.052 & 0.068 & 0.074 & 0.070 & 0.966 & 0.955 \\
				&  AN IV  & 0.002 & 0.055 & 0.070 & 0.070 & 0.067 & 0.952 & 0.930 \\
				\hline\\[-2ex]
				120  & UN  & -0.001 & 0.062 & 0.070 & 0.078 & 0.076 & 0.967 & 0.964 \\
				&  AN I  & -0.001 & 0.040 & 0.053 & 0.062 & 0.061 & 0.977 & 0.973 \\
				&  AN II  & -0.001 & 0.044 & 0.056 & 0.064 & 0.062 & 0.977 & 0.974 \\
				&  AN III  & 0.001 & 0.038 & 0.051 & 0.052 & 0.051 & 0.938 & 0.926 \\
				&  AN IV  & 0.001 & 0.039 & 0.052 & 0.051 & 0.050 & 0.931 & 0.925 \\
				\hline
			\end{tabular}
	\end{center}
\end{table}

In comparison, for the unadjusted, ANCOVA I and II estimators, the CRSE approach yields ASEs closer to respective ESEs even when $I$ is relatively small, but it may grow conservative when $I$ increases. For ANCOVA III and IV, the CRSE approach can underestimate the variance when $I$ is relatively small, but as $I$ increases, the yielded ASEs grow to approach respective ESEs. Overall, when the number of clusters is limited, the CRSE approach demonstrates better finite-sample properties for the unadjusted, ANCOVA I and II estimators, while the DB approach demonstrates better finite-sample properties for the ANCOVA III and IV estimators.

\begin{table}[htbp] 
	\caption{Results for the cell-average treatment effect ($\tau^{cell}$) from simulation study I comparing performances of five estimators, UN = unadjusted, AN I-IV = ANCOVA I-IV. Evaluation metrics: BIAS = relative bias; RMSE = root mean squared error; ESE = empirical standard error; ASE = average standard error, where, DB = standard errors via the design-based plug-in estimator, and CRSE = cluster-robust standard errors; Coverage = empirical coverage of 95\% confidence intervals over 1,000 simulation replications.}\label{tab:sim-1-ate-3}
	\par
	\begin{center}
			\begin{tabular}{cc r cc cc cc} 
				\hline 
				& & & & & \multicolumn{2}{c}{ASE} & \multicolumn{2}{c}{Coverage} \\
				\cline{6-9} \\[-1.5ex]
				$I$ & Estimator & \multicolumn{1}{c}{BIAS} & RMSE & ESE & DB & CRSE & DB & CRSE \\
				\hline\\[-2ex]
				18 & UN  & -0.012 & 0.149 & 0.171 & 0.210 & 0.171 & 0.963 & 0.923 \\
				& AN I  & 0.005 & 0.096 & 0.126 & 0.165 & 0.135 & 0.980 & 0.954 \\
				& AN II  & 0.015 & 0.109 & 0.137 & 0.162 & 0.134 & 0.967 & 0.930 \\
				& AN III  & 0.004 & 0.092 & 0.124 & 0.135 & 0.111 & 0.954 & 0.917 \\
				& AN IV  & 0.030 & 0.103 & 0.132 & 0.115 & 0.092 & 0.896 & 0.808 \\
				\hline\\[-2ex]
				30 & UN  & -0.014 & 0.119 & 0.134 & 0.157 & 0.140 & 0.971 & 0.946 \\
				& AN I  & 0.001 & 0.078 & 0.101 & 0.125 & 0.111 & 0.974 & 0.963 \\
				& AN II  & 0.004 & 0.088 & 0.110 & 0.125 & 0.113 & 0.964 & 0.948 \\
				& AN III  & 0.002 & 0.074 & 0.098 & 0.103 & 0.093 & 0.951 & 0.934 \\
				& AN IV  & 0.014 & 0.081 & 0.103 & 0.094 & 0.082 & 0.917 & 0.872 \\
				\hline\\[-2ex]
				60 & UN & -0.001 & 0.079 & 0.089 & 0.109 & 0.103 & 0.980 & 0.975 \\
				& AN I  & 0.005 & 0.052 & 0.066 & 0.088 & 0.083 & 0.986 & 0.984 \\
				& AN II  & 0.008 & 0.057 & 0.070 & 0.090 & 0.085 & 0.983 & 0.978 \\
				& AN III  & 0.005 & 0.050 & 0.066 & 0.073 & 0.069 & 0.965 & 0.958 \\
				& AN IV  & 0.012 & 0.052 & 0.067 & 0.070 & 0.066 & 0.953 & 0.942 \\
				\hline\\[-2ex]
				120 & UN  & 0.001 & 0.058 & 0.066 & 0.076 & 0.074 & 0.969 & 0.967 \\
				& AN I  & 0.003 & 0.039 & 0.051 & 0.061 & 0.060 & 0.975 & 0.970 \\
				& AN II  & 0.003 & 0.043 & 0.054 & 0.063 & 0.061 & 0.972 & 0.967 \\
				& AN III  & 0.004 & 0.037 & 0.049 & 0.051 & 0.050 & 0.947 & 0.940 \\
				& AN IV  & 0.005 & 0.038 & 0.050 & 0.050 & 0.049 & 0.940 & 0.935 \\
				\hline
			\end{tabular}
	\end{center}
\end{table}

Under the parallel-arm cluster randomized experiments, \citet{Middleton2015} and \citet{Su2021} showed that when using the regression based on the scaled cluster totals, including the cluster size as an additional covariate can lead to asymptotic efficiency gain. Although we are unable to provide an analogous theoretical result for the SW-CREs with individual-level data regression models, we repeat the above simulations by including $X_{ij3}=N_{ij}$ as an additional covariate to empirically investigate this issue. The results are presented in Web Tables 8-10 of Web Appendix D. When the number of clusters is sufficiently large, we observe that adjusting for $N_{ij}$ can improve the estimation precision and lead to a smaller ESE for all ANCOVA estimators. However, when only a limited number of clusters are included, adjusting for $N_{ij}$ an additional covariate results in finite-sample efficiency loss for ANCOVA II and IV estimators. This may be explained by the fact that ANCOVA II and IV would require estimating a much larger number of parameters to accommodate an additional covariate than ANCOVA I and III. Otherwise, the performances of the variance estimators are generally similar to our main simulations without including $N_{ij}$.

\subsection{Simulation study II}

\begin{figure}[htbp]
	\centering
	\includegraphics[width=0.8\textwidth]{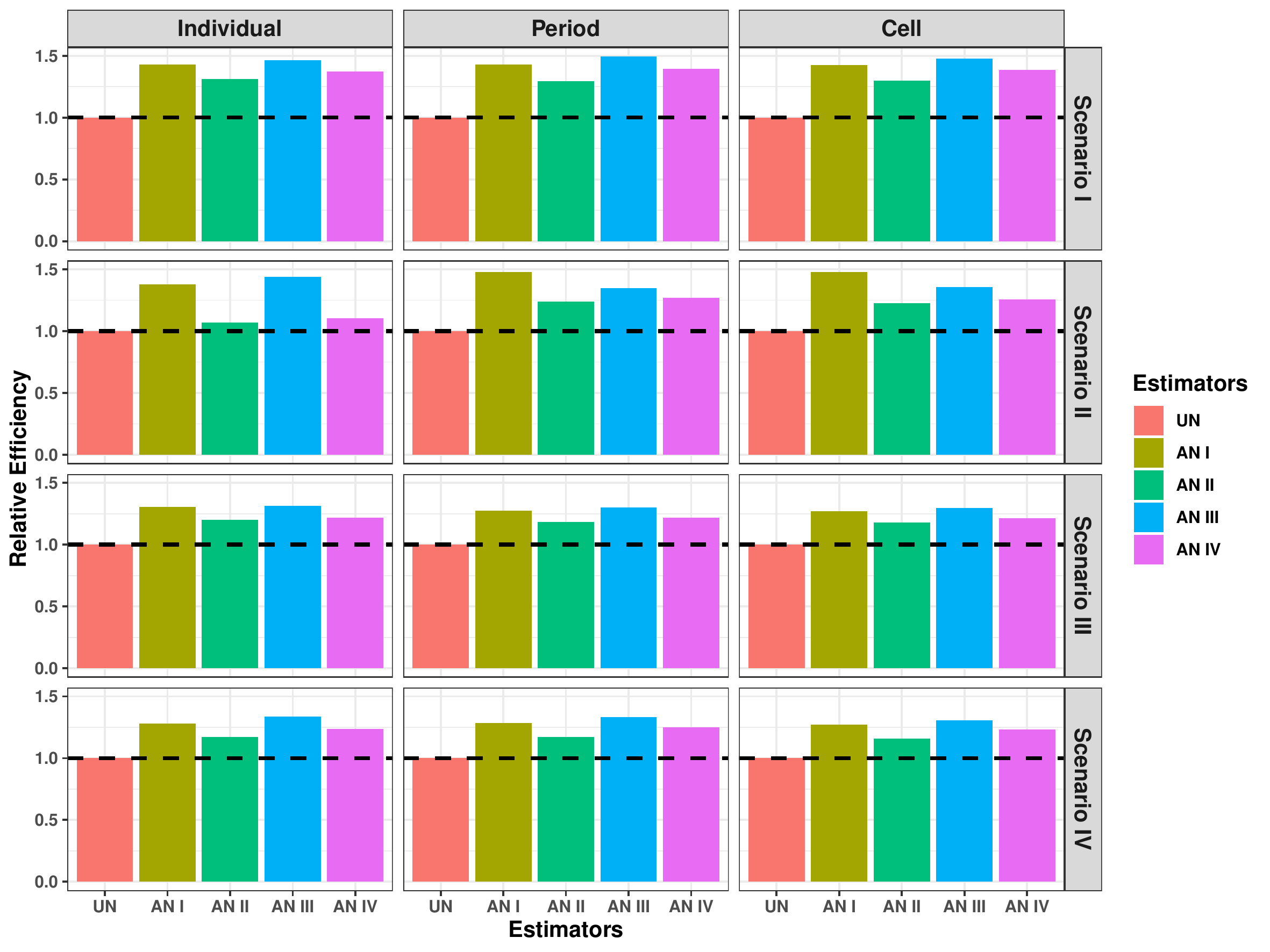}
	\caption{Relative efficiency of proposed estimators from simulation study II with 1,000 simulation replications. UN = unadjusted, AN I-IV = ANCOVA I-IV. The number of clusters $I=18$.}
	\label{fig:sim-2-18}
\end{figure}

\begin{figure}[htbp]
	\centering
	\includegraphics[width=0.8\textwidth]{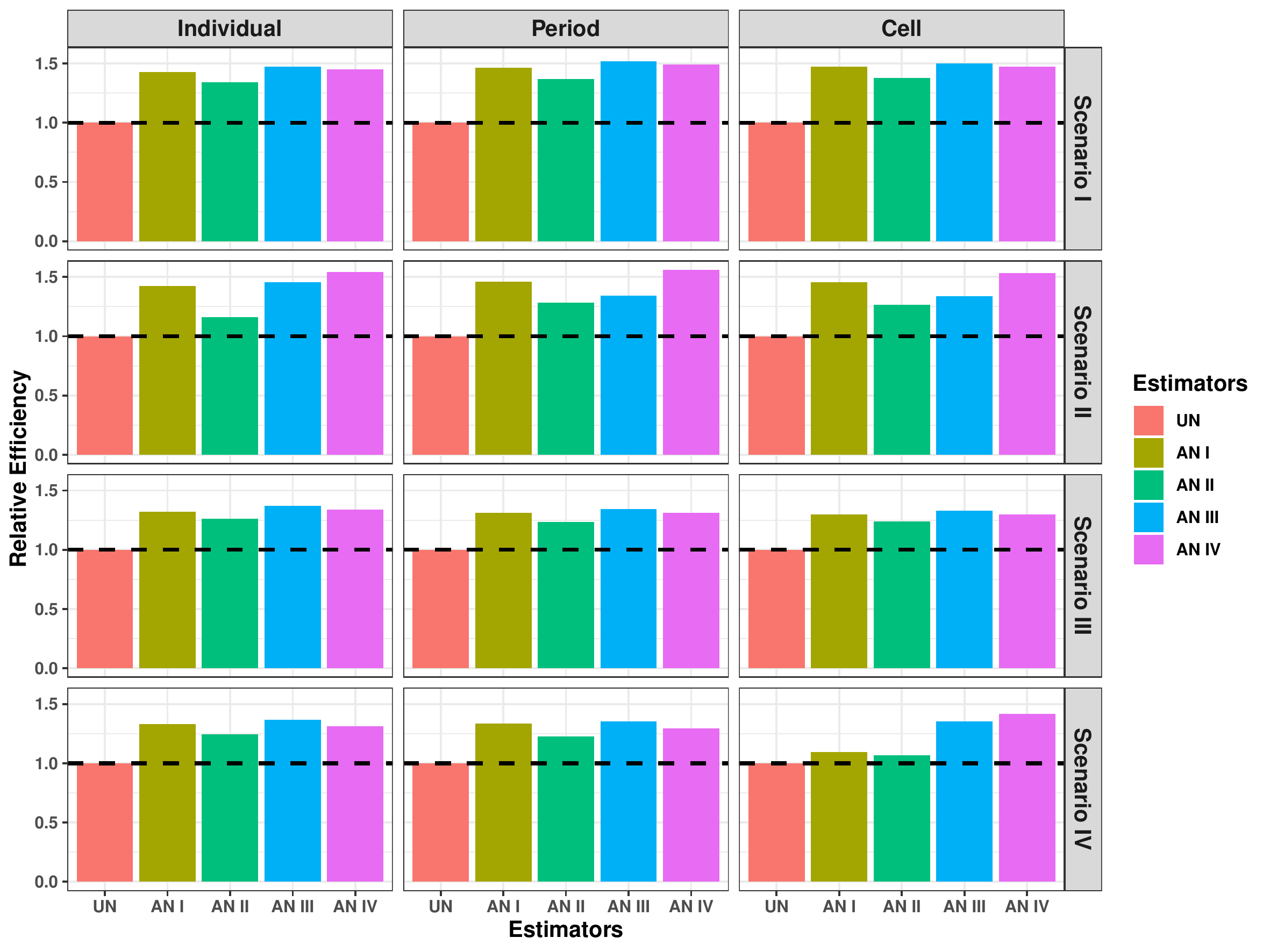}
	\caption{Relative efficiency of proposed estimators from simulation study II with 1,000 simulation replications. UN = unadjusted, AN I-IV = ANCOVA I-IV. The number of clusters $I=60$.}
	\label{fig:sim-2-60}
\end{figure}

In simulation study II, we compare the performances of model-assisted estimators in terms of their relative efficiency under several carefully chosen data-generating processes to further inform the choices of working models. The same randomization scheme in simulation study I is maintained, and the same covariate-generating processes for $X_{ij1}$ and $X_{ijk2}$ are considered. However, we additionally consider the following four additional scenarios to simulate the potential outcomes:
\begin{enumerate}
	\item Scenario I: potential outcomes are not affected by the cluster-period size, where
	\begin{align*}
		Y_{ijk}(0) &=\frac{j+1}{J+2}+X_{ij1}+\left(X_{ijk2}-\overline{X}_{j2}\right)^2+c_i+e_{ijk},\\
		Y_{ijk}(1) &=Y_{ijk}(0)+0.5X_{ij1}+\left(X_{ijk2}-\overline{X}_{j2}\right)^3,
	\end{align*}
	with $c_i\sim\calN(0,0.1)$ and $e_{ijk}\sim\calN(0,0.9)$. In this case, the three estimands have the same value. This type of data-generating process, where the cluster-period size is uninformative, has been the most typical in the existing literature for SW-CREs \citep{Li2021}.
	\item Scenario II: the true individual treatment effect depends on the cluster-period size and further the calendar period through the period-specific main covariate effect in $Y_{ijk}$(1), where
	\begin{align*}
		Y_{ijk}(0) &=\frac{j+1}{J+2}+X_{ij1}+\left(X_{ijk2}-\overline{X}_{j2}\right)^2+c_i+e_{ijk},\\
		Y_{ijk}(1) &=Y_{ijk}(0)+\frac{2N_{ij}I}{(J+2)^{-1}\sum_{j=0}^{J+1}N_j}+0.5(j+1)X_{ij1}+\frac{j+1}{J+2}\left(X_{ijk2}-\overline{X}_{j2}\right)^3,
	\end{align*}
	with $c_i\sim\calN(0,0.1)$ and $e_{ijk}\sim\calN(0,0.9)$. This type of data-generating process may be plausible in the presence of seasonality of treatment effects or when there is a background secular trend (e.g., another concurrent intervention) that can interact with the treatment of interest.
	\item Scenario III: the random effects are more complicated than the simple compound symmetry structure, where
	\begin{align*}
		Y_{ijk}(0) &=\frac{j+1}{J+2}+X_{ij1}+\left(X_{ijk2}-\overline{X}_{j2}\right)^2+c_i+b_{ij}+e_{ijk},\\
		Y_{ijk}(1) &=Y_{ijk}(0)+\frac{2N_{ij}I}{(J+2)^{-1}\sum_{j=0}^{J+1}N_j}+0.5X_{ij1}+\left(X_{ijk2}-\overline{X}_{j2}\right)^3 + d_i,
	\end{align*}
	with $c_i\sim\calN(0,0.05)$, $b_{ij}\sim\calN(0,0.05)$, $d_i\sim\calN(0,0.1)$, and $e_{ijk}\sim\calN(0,0.9)$. Cluster-period-specific random effects, $b_{ij}$, and cluster random effects, $d_i$, are independent and independent from other components. In the SW-CRE literature, this correlation structure can be viewed as a combination of the nested exchangeable structure \citep{hooper2016sample} with a random intervention effect. 
	\item Scenario IV: the distributions of random effects are non-normal and skewed, where
	\begin{align*}
		Y_{ijk}(0) &=\frac{j+1}{J+2}+X_{ij1}+\left(X_{ijk2}-\overline{X}_{j2}\right)^2+c_i+e_{ijk},\\
		Y_{ijk}(1) &=Y_{ijk}(0)+\frac{2N_{ij}I}{(J+2)^{-1}\sum_{j=0}^{J+1}N_j}+0.5X_{ij1}+\left(X_{ijk2}-\overline{X}_{j2}\right)^3,
	\end{align*}
	with $c_i\sim\calC\calG(1,0.1)$, a centered gamma distribution with variance 0.1, and $e_{ijk}\sim\calC\calP(0.9)$, a centered Poisson distribution with variance 0.9. 
\end{enumerate}
We consider two cases where the sample size is relatively small ($I=18$) and large ($I=60$). Same as simulation study I, we conduct 1,000 simulation replications with models \eqref{eq:gen-model}-\eqref{eq:int-model-tv} fitted with $X_{ij1}$, $X_{ijk2}$ but without further adjusting for cluster-period size; results for all metrics (BIAS, RMSE, ESE, ASE) and the empirical coverage percentages of 95\% confidence intervals (CIs) using the DB and CRSE estimators are analogously obtained. Here, we focus on presenting the relative efficiency (RE) of each ANCOVA estimator to the unadjusted estimator in Figure \ref{fig:sim-2-18} and \ref{fig:sim-2-60} (a larger RE indicates that the ANCOVA estimator is more efficient in finite samples); more detailed results are available in Web Tables 2-7 of Web Appendix C. 

As expected, the unadjusted estimator is less efficient than the ANCOVA estimators, confirming the benefits of covariate adjustment under the staggered cluster rollout design. When the number of clusters is small ($I=18$), ANCOVA I and III exhibit higher RE than the other estimators. Comparing the four ANCOVA estimators, estimators not including period-specific effects (ANCOVA I and III) demonstrate higher RE, likely because the ANCOVA II and IV estimators require estimating a significantly larger number of model parameters, which can lead to finite-sample efficiency loss. Comparing the two pairs of ANCOVA estimators, i.e., ANCOVA I vs. III and II vs. IV, we found that the estimator, including interactions between treatment indicators and covariates, can improve RE in most settings, except for estimating $\tau^{period}$ and $\tau^{cell}$ under scenario II, where ANCOVA I has higher RE than ANCOVA III. This slightly contradicts the earlier results found for individually randomized experiments \citep{Lin2013}, where the fully-interacted model is always asymptotically more efficient. Similar observations were discussed in \citet{Su2021} for parallel-arm cluster randomized experiments, where the fully-interacted individual-level regression estimator does not necessarily improve the asymptotic efficiency over its counterpart without the treatment-by-covariate interactions. With more clusters ($I=60$), Figure \ref{fig:sim-2-60} shows that estimators, including treatment-by-covariate interactions, outperform their counterparts without interactions in most simulation settings. In particular, ANCOVA IV, the estimator assisted by the most richly parameterized working model, yields the highest RE in Scenario II, where the true individual treatment effect depends on the period through the period-specific main covariate effect in $Y_{ijk}(1)$. This is likely because ANCOVA IV most accurately reflects the true data-generating process where period-specific covariate effects and treatment-by-covariate interactions are simultaneously accounted for by the model structure (and such structures are estimable with sufficient data from 60 clusters). Overall, if sufficient clusters are present, we recommend ANCOVA III and IV for efficiency considerations. If only a limited number of clusters are present, using ANCOVA III does not appear to compromise finite-sample efficiency under a relatively parsimonious working model parameterization and may be preferred.

Finally, in Web Figures 2-3 of Web Appendix D, we present the relative efficiency results when $N_{ij}$ is included as an additional covariate in the ANCOVA working models. (More detailed results are available in Web Tables 11-16 of Web Appendix D.) While the overall observations are similar to the above simulations without adjusting for $N_{ij}$, a notable difference is that with $I=18$, ANCOVA IV becomes the least efficient and can even be substantially less efficient than the unadjusted estimator (occasionally the $\text{RE}=0.5$). This suggests that one should caution the adjustment for cluster-period size via ANCOVA IV when a limited number of clusters cannot support the large number of parameters that need to be estimated for this working model.

\section{Application to the Washington State Expedited Partner Therapy study} \label{sec:application}

We illustrate the application of model-assisted estimators by analyzing the Washington State Expedited Partner Therapy (EPT) study. The Washington State EPT study is an SW-CRE designed to evaluate the effectiveness of an expedited patient-delivered partner notification strategy, which aims to treat the sex partners of persons with sexually transmitted infections without their medical evaluation and thus increase partner treatment and decrease gonorrhea and chlamydia reinfection rates \citep{Golden2015}. The trial was conducted from October 2007 to August 2009, with four waves (1 pre-rollout period, 3 rollout periods, and 1 post-rollout period) separated by six-month intervals. A total of 22 local health jurisdictions (LHJs, i.e., clusters) were randomly assigned to the four different treatment adoption times and provided individual-level outcome data that was measured based on distinct sentinel women sampled during each period. In each of the three rollout periods, six previously untreated LHJs were given the intervention, and in the post-rollout period, the remaining four untreated LHJs were assigned to the treatment arm. We focus on the binary outcome, Chlamydia infection status, in this analysis, with a value equal to 1 if the sentinel woman reports Chlamydia at the time of assessment and 0 otherwise; therefore, all estimands are interpreted on the risk difference scale. The cluster-period (cell) size $N_{ij}$ during rollout ranges from 41 to 1553, with a standard deviation of 331 (see Figure 1 in \citet{Li2021marginal} and \citet{Tian2022} for a detailed depiction of cluster-period sizes over the pre-rollout, rollout, and post-rollout periods). Due to the substantial variation in cluster sizes, informative cluster size may not be ruled out, and we are interested in quantifying the three different treatment effect estimands defined in Table \ref{tab:estimands}.

We apply the undjusted estimator and the proposed model-assisted ANCOVA estimators (ANCOVA I-IV) to the Washington State EPT study data by fitting models \eqref{eq:gen-model}-\eqref{eq:int-model-tv}, with two covariates---age measured at baseline, and an LHJ-level Chlamydia prevalence measured at baseline---that are believed to have good prognostic values. Estimation results from various estimators with standard errors obtained from two different approaches are given in Table \ref{tab:ept}. Results adjusted for cluster-period size as an additional covariate are presented in Web Table 17 of Web Appendix D and are generally similar.

\begin{table}[ht] 
	\caption{Estimated average treatment effects on the risk difference scale and their standard errors for the Washington State EPT study. Standard errors are given in parentheses. DB = design-based standard error. CRSE = cluster-robust standard error.}\label{tab:ept}
	\par
	\begin{center}
			\begin{tabular}{l ll c ll c ll} 
				\hline 
				& \multicolumn{2}{c}{$\tau^{ind}$} && \multicolumn{2}{c}{$\tau^{period}$} && \multicolumn{2}{c}{$\tau^{cell}$} \\
				\cline{2-3} \cline{5-6} \cline{8-9} \\[-1.5ex]
				Estimators & \multicolumn{1}{c}{DB} & \multicolumn{1}{c}{CRSE} && \multicolumn{1}{c}{DB} & \multicolumn{1}{c}{CRSE} && \multicolumn{1}{c}{DB} & \multicolumn{1}{c}{CRSE} \\
				\hline\\[-2ex]
				Unadjusted & -0.0032 & -0.0032 && -0.0065 & -0.0065 && -0.0137 & -0.0137 \\
				& (0.0049) & (0.0044) && (0.0053) & (0.0049) && (0.0102) & (0.0093) \\[1.5ex]
				ANCOVA I & -0.0068 & -0.0068$^*$ && -0.0095$^*$ & -0.0095$^*$ && -0.0140 & -0.0140 \\
				& (0.0045) & (0.0040) && (0.0049) & (0.0043) && (0.0096) & (0.0087) \\[1.5ex]
				ANCOVA II & -0.0066 & -0.0066$^*$ && -0.0092$^*$ & -0.0092$^*$ && -0.0143 & -0.0143 \\
				& (0.0044) & (0.0040) && (0.0048) & (0.0043) && (0.0096) & (0.0087) \\[1.5ex]
				ANCOVA III & -0.0087$^*$ & -0.0087$^*$ && -0.0111$^{**}$ & -0.0111$^{**}$ && -0.0140 & -0.0140 \\
				& (0.0045) & (0.0043) && (0.0044) & (0.0041) && (0.0095) & (0.0087) \\[1.5ex]
				ANCOVA IV & -0.0037 & -0.0037 && -0.0071$^*$ & -0.0071 && -0.0142 & -0.0142 \\
				& (0.0039) & (0.0049) && (0.0040) & (0.0045) && (0.0094) & (0.0090) \\
				\hline
				\multicolumn{9}{l}{$^{**}$\footnotesize Statistically significant at the 5\% level, two-tailed test.}\\
				\multicolumn{9}{l}{~$^*$\footnotesize Statistically significant at the 10\% level, two-tailed test.}
			\end{tabular}
	\end{center}
\end{table}

Overall, the results indicate that the intervention implemented in the Washington State EPT study reduces the likelihood of Chlamydia, as the signs of all point estimates are negative. The estimates have a clear ordering such that $|\widehat{\tau}^{cell}|>|\widehat{\tau}^{period}|>|\widehat{\tau}^{ind}|$, suggesting the largest effect may be observed once we consider an average at the cell level. For each of the three estimands, the unadjusted estimator yields the largest standard error estimate, and the ANCOVA I-IV estimators appear to improve the estimation precision with a smaller standard error estimate. Our simulations show that the ANCOVA III estimator demonstrates the highest estimation efficiency under most scenarios, and the DB approach provides a relatively accurate estimation of the uncertainty for this approach; therefore, we primarily interpret this result here. For the individual-average treatment effect $\tau^{ind}$, the point estimate is $-0.0087$; therefore, after averaging over all sentinel women during rollout, the EPT intervention can prevent 87 positive Chlamydia infection cases per 10 thousand women during a six-month interval (with statistical significance at the 10\% level). For the period-average treatment effect $\tau^{period}$, the point estimate is $-0.0111$; therefore, during an average rollout period, after averaging over all sentinel women, the EPT intervention can prevent 111 positive Chlamydia infection cases per 10 thousand women (with statistical significance at the 5\% level). For the cell-average treatment effect ($\tau^{cell}$), the point estimate is $-0.0142$; therefore, after averaging all cluster-period cells during rollout, the EPT intervention can prevent 142 positive Chlamydia infection cases per 10 thousand women. These results generally corroborate the model-based analysis results in \citet{Golden2015} and \citet{Li2021marginal}, even though the previous analyses focused on the ratio estimands and did not address variations in the treatment effect estimands due to potentially informative cluster-period size.

\section{Discussion} \label{sec:discussion}

In this article, we have studied model-assisted analyses of SW-CREs and elucidated considerations on treatment effect estimands, covariate adjustment, and statistical inference strategies. Specifically, we examined a class of weighted average treatment effects as nonparametric estimands, where three interpretable members can be obtained by selecting appropriate weights. Leveraging the ANCOVA working models, a class of estimators was developed to exploit baseline covariates information and improve estimation efficiency over the conventional unadjusted difference-in-means estimators. Asymptotic results for proposed estimators were established via finite population Central Limit Theorems, i.e., results that confirm that the proposed estimators are consistent with model-free estimands and asymptotically normal as the number of clusters increases to infinity. We have also conducted a series of simulation studies to evaluate the finite-sample properties of the ANCOVA estimators and compare their performances to inform practical recommendations. To the best of our knowledge, this is the first effort to systematically investigate treatment effect estimands and model-robust causal inference methods for SW-CREs that analyze individual-level data.

While it is generally challenging to analytically compare the four ANCOVA models under the staggered rollout randomization designs, we seek to conduct numerical experiments to inform model choices in practice. The following main messages are generated from our simulation evaluations. (1) All five estimators, including the unadjusted estimator, have negligible biases in finite samples, corroborating our theoretical findings, and furthermore, as the number of clusters increases, the relative biases decrease. This indicates that including covariates in the ANCOVA models does not compromise bias, as long as the correct weights are specified to target a specific estimand. (2) The DB variance estimator can be conservative for the unadjusted, ANCOVA I and II estimators. While the DB approach demonstrates the desired estimation accuracy for the ANCOVA III estimator regardless of sample size, it often underestimates the variance of the ANCOVA IV estimator when the sample size is small. (3) The CRSE variance estimator yields accurate estimates for the unadjusted, ANCOVA I and II estimators with limited clusters and can be conservative in larger samples. For ANCOVA III and IV, the CRSE estimator tends to underestimate the true variance, though the degree of underestimation diminishes when the sample size increases. (4) The unadjusted estimator generally exhibits the lowest RE compared to ANCOVA, which aligns with the result from \citet{Lin2013}, \citet{Su2021} and \citet{wang2022model} that adjusting for covariates increases estimation efficiency in individually randomized experiments and parallel-arm cluster randomized experiments. (5) Comparing the estimators assisted by models, including treatment-by-covariate interactions with their counterparts without interactions, the fully-interacted estimators show higher RE in most settings. However, the model without interactions performs better in a few simulation situations. This finding aligns with results in \citet{Su2021} that including interactions in analyzing individual-level data does not always increase estimation efficiency in parallel-arm cluster randomized experiments. (6) With a limited number of clusters, estimators assisted by more parsimonious models, i.e., ANCOVA I and III, tend to give better performances in terms of estimation efficiency. When the sample size increases, estimators assisted by more richly parameterized models, especially ANCOVA IV, demonstrate higher estimation efficiency. (7) Adjusting for cluster-period size as an additional covariate in ANCOVA models can improve estimation efficiency when the number of clusters is large. With a small number of clusters, however, including cluster-period size may lead to a less efficient estimator under ANCOVA IV than the unadjusted estimator. Summarizing this empirical evidence, we find that if the number of clusters is limited, ANCOVA III coupled with the DB variance estimator can be an adequate strategy for statistical inference.

There has been substantial interest in studying optimal variance estimation of the treatment effect estimator in the current literature for SW-CREs. Still, these efforts are confined to model-based inference under the super population perspective. We take an alternative approach to define model-free treatment effect estimands under a finite population perspective and then develop model-assisted estimators that explicitly target the model-free estimands; unlike the super population perspective, we consider the randomness of the estimator to arise solely from the staggered cluster randomization regime rather than the sampling of clusters from an infinite super population. For uncertainty quantification under a finite population perspective, we first develop a design-based variance (or equivalently standard error) estimator that is at most conservative for the true randomization-based variance in Theorem \ref{thm:1}. This observation is borne out by our simulation results, where the DB standard error estimator often overestimates the true Monte Carlo standard error with a small number of clusters. In addition, we have also considered the CRSE estimator, originally developed for generalized estimating equations under a super population perspective. This variance estimator is of natural interest because our ANCOVA estimators are all special cases of the weighted generalized estimating equations under working independence \citep{Liang1986,preisser2002performance}, and software for generalized estimating equations is widely accessible. In individual randomized experiments, \citet{ding2017bridging} have established insightful connections between the variance estimators developed under the finite population and super population perspectives based on the difference-in-means estimator. In the context of parallel-arm cluster randomized experiments, \citet{Su2021} have formally proved that the CRSE estimator is a conservative approximation to the true randomization-based standard error for the unadjusted and two ANCOVA estimators with individual-level data. Obtaining similar results in our setting, however, is challenging because (1) the design matrix associated with our estimators is substantially more complex than its counterpart under a parallel-arm design, and (2) the staggered randomization scheme introduces difficulty in connecting the CRSE estimator and the true randomization-based standard error; we thus leave this investigation to future work and only compare the different variance estimators via simulations. In simulations, the CRSE estimates are, on average, smaller than the DB standard error estimates and better approximates the Monte Carlo standard error, especially when the number of clusters is limited. Moreover, regardless of estimands, the CRSE estimates are frequently larger than the Monte Carlo standard error when the working model is not over-parameterized (e.g., ANCOVA I and II), especially when the number of clusters increases. This aligns with the theoretical findings in \citet{Su2021} for parallel-arm designs. When fitting ANCOVA III and IV with a limited number of clusters, the CRSE estimates can often, on average, be smaller than the Monte Carlo standard error, which is reminiscent of previous findings for model-based inference in stepped wedge designs, where the CRSE estimator typically underestimates the Monte Carlo standard error in small samples. Therefore, various bias-corrected CRSE estimators have been studied to improve small-sample model-based inference in stepped wedge designs \citep{li2018sample}. For model-assisted inference in stepped wedge designs, our results indicate that the bias-corrected CRSE estimator may not be necessary for working ANCOVA models without the treatment-by-covariate interactions but may remain useful to over-parameterized working models, such as ANCOVA IV. This is a topic of future research, and bias-corrected CRSE for ANCOVA IV may be directly adapted from those developed for weighted generalized estimating equations (e.g., Web Appendix 3 in \citet{turner2020properties}).

Our current study has several potential limitations. First, we have focused on estimators assisted by ANCOVA models without the duration effects. When the true treatment effect depends on the length of exposure, \citet{kenny2021analysis} and \citet{maleyeff2022assessing} showed that the model assuming a constant treatment effect can lead to an effect with an opposite direction from the true effect. Similar observations have also been discussed in \citet{Sun2021} for dynamic difference-in-differences. It may be possible to modify our ANCOVA working models to target a class of duration-specific weighted average treatment effects, and we leave this non-trivial development under the finite population perspective to future work. Second, our finite population framework and estimands currently do not explicitly address the missing potential outcomes during the pre-and post-rollout periods, as treatment positivity is technically violated by the construction of the stepped wedge design in those periods. It may be intriguing to expand our estimand definitions to accommodate those periods, but identification would necessarily require additional assumptions (either structural or modeling) to extrapolate from the rollout periods to the pre-and post-rollout periods. For example, the conventional linear mixed models \citep{Li2021} consider random effects at the cluster and cluster-period levels to implicitly impute the unobserved potential outcomes during all study periods and have been demonstrated to be model-robust under certain structural assumptions for parallel-arm cluster randomized experiments \citep{wang2021mixed,wang2022model}. Under stepped wedge designs and focusing on the cluster-average treatment effect estimand, \citet{wang2024achieve} has also developed model-assisted estimators based on linear mixed models and generalized estimating equation under several treatment effect structures, and similarly suggested the use of the cluster-robust sandwich variance estimator for valid inference. Despite the technical differences between the finite population and super population perspectives, in future work, it will be interesting to empirically assess whether the \citet{wang2024achieve} estimators can further improve efficiency over our current ANCOVA estimators assuming working independence. Finally, although we operate under a finite population perspective, our main asymptotic theory still assumes the number of clusters to approach infinity with bounded cluster sizes. It is intriguing to develop complementary asymptotic theory under an alternative scheme where the number of clusters is fixed, but the cluster size approaches infinity. \citet{xie2003asymptotics} studied the asymptotic results of generalized estimating equation estimators when either the number of clusters or the cluster sizes or both approach infinity, and it may be useful to extend their results to inform causal inference with cluster randomized experiments under alternative asymptotic schemes.

\bmsection*{Acknowledgments}
The authors would like to thank the editor, the associate editor, and four anonymous reviewers for their valuable input that has significantly improved the quality of this paper. The authors also would like to thank Professor James P. Hughes for sharing the data from the Washington State Expedited Partner Therapy (EPT) study.

\bmsection*{Financial disclosure}

Research in this article was supported by a Patient-Centered Outcomes Research Institute Award\textsuperscript{\textregistered} (PCORI\textsuperscript{\textregistered} Award ME-2022C2-27676). The statements presented in this article are solely the responsibility of the authors and do not necessarily represent the official views of PCORI\textsuperscript{\textregistered}, its Board of Governors, or the Methodology Committee. 

\bmsection*{Conflict of interest}

The authors declare no potential conflict of interests.

\bmsection*{Supporting information}

Proofs and additional simulation results are available in the Web Appendix.

\bibliography{MA-SWD}

\end{document}